\ifx\pdfoutput\undefined\else\pdfoutput=1\fi  
\documentclass[preprint,12pt,authoryear]{elsarticle}

\usepackage[utf8]{inputenc}
\usepackage{amsmath,amssymb}
\usepackage{graphicx}
\usepackage{booktabs}
\usepackage{array}
\usepackage{url}

\journal{Journal of Computational Physics}

\begin{document}

\begin{frontmatter}

\title{When a neural surrogate cannot accelerate a solver:
runtime share, closed-loop drift, and the economics of
uncertainty gating in a stiff coupled simulation}

\author[ethz]{L.~Th\"ummler\corref{cor1}}
\ead{lthuemmler@ethz.ch}
\cortext[cor1]{Corresponding author.}
\author[aei]{T.~Kuroda}
\ead{takami.kuroda@aei.mpg.de}
\address[ethz]{Department of Mathematics, ETH Z\"urich, R\"amistrasse 101,
8092 Z\"urich, Switzerland}
\address[aei]{Max Planck Institute for Gravitational Physics (Albert
Einstein Institute), Am M\"uhlenberg 1, D-14476 Potsdam, Germany}

\begin{abstract}
Learned surrogates for expensive inner solver blocks are a widely pursued
route to faster multiphysics simulation. We report a controlled, end-to-end
negative result and identify three structural barriers, none of them a
deficiency of the network we trained. The testbed is the implicit Newton
solve coupling energy-dependent neutrino radiation to matter in a
general-relativistic radiation-hydrodynamics code, its most expensive
physics routine \emph{per call}.

First, per-call cost and share of runtime are different quantities, and only
the second bounds acceleration. An exclusive self-time profile puts the
target block at $16.9\%$ of critical-rank wall clock, capping any surrogate
at $\approx\!1.2\times$ by Amdahl's law. A surrogate $5.8\times$ cheaper per
call merely ties the solver, and the configuration stable enough to run
without fallback reaches only parity. Second, offline accuracy cannot rank surrogates
for deployment: across fourteen networks the pooled Spearman
error-versus-survival correlation ($\rho=+0.73$) is a between-family
confound that vanishes under control ($\rho=-0.04$). Third, a \emph{correct} out-of-distribution gate cannot
accelerate a loop that leaves its training distribution. We give the
break-even deferral fraction in closed form: because the visited states sit
$73\times$ off the data manifold, the gate defers $96.8$ to $99.7\%$ of
cells, almost invariant to surrogate quality. Including its own cost, the
gated loop is a $0.94$ to $0.96\times$ slowdown.

We further separate stability from fidelity: a never-crashing gated run
accumulates a linear $-19.9\%$ density bias over $6000$ steps. The error is
a \emph{directed}, ballistically accumulating bias, not the variance-driven
divergence the autoregressive literature targets.
\end{abstract}


\end{frontmatter}

\section{Introduction}\label{sec:intro}

A recurring proposal in computational physics is to replace an expensive
inner solver block, a stiff ordinary differential equation integrator, a
root find, a per-cell implicit coupling, with a learned surrogate whose
forward pass costs microseconds. The appeal is immediate and the offline
evidence is usually encouraging: such surrogates routinely reach
coefficients of determination above $0.99$ on held-out data and per-call
speedups of one to two orders of magnitude.

Whether that translates into a faster simulation is a separate question, and
it is answered by three quantities that offline benchmarks do not measure.
The first is the block's \emph{share of runtime}, which bounds the
achievable end-to-end gain by Amdahl's law \citep{amdahl1967} regardless of
surrogate quality.
The second is the surrogate's behaviour \emph{in the closed loop}, where its
outputs become the next step's inputs and errors compound, a regime in which
per-step accuracy is a poor predictor of anything. The third arises only if
one attempts the natural fix for the second, an uncertainty gate that routes
uncertain cells back to the exact solver: such a gate has its own cost and
its own deferral fraction, and these determine whether gating can pay at all.

This paper works all three through in a single system, with end-to-end wall
clock as the metric throughout. Our testbed is the implicit neutrino-matter
coupling solve in a one-dimensional general-relativistic
radiation-hydrodynamics code. We chose it because it is an unusually
favourable target on the criterion that motivates most such work: it is by a
wide margin the most expensive physics routine \emph{per call} in that code,
$3.4\times10^{-4}\,$s per cell per step, and it is called once per cell per
step. If a learned surrogate can accelerate any block, it should accelerate
this one.

It does not, and the reasons generalise beyond the testbed. We state our
contributions explicitly.

\begin{enumerate}
\item \textbf{A measured acceleration budget, and the distinction it
enforces} (Sec.~\ref{sec:budget}). We profile the complete time step with
exclusive self-time timers and validate the decomposition by summation. The
target routine is $10.5\%$ of mean wall clock and $16.9\%$ on the critical
rank, capping any surrogate at $\approx\!1.2\times$. We trace how a per-call
cost figure in our own earlier planning was mistaken for a runtime share,
and we survey the literature for cases where replacing a minority-cost block
produced a measured end-to-end speedup, finding none.

\item \textbf{Evidence that offline metrics cannot select surrogates}
(Sec.~\ref{sec:offline}). Fourteen networks on a common held-out set, with
the pooled correlation shown to be a between-family confound, plus a seed
study establishing that single-run survival counts are not comparable.

\item \textbf{Closed-form gate economics, with measured inputs}
(Sec.~\ref{sec:gate}). We derive the break-even deferral fraction, measure
the gate cost in situ at three implementation qualities, and show the
measured deferral is nearly invariant to surrogate quality, which places the
operating point beyond break-even structurally rather than incidentally.

\item \textbf{A separation of stability from fidelity}
(Sec.~\ref{sec:fidelity}). A gated run that never crashes still accumulates
a linear density bias to $-19.9\%$ over $6000$ steps.

\item \textbf{A mechanism reframing} (Sec.~\ref{sec:mechanism}). The
closed-loop failure here is driven by a \emph{directed} bias accumulating
ballistically, not the variance-driven blow-up that noise injection and
pushforward losses are designed to damp, which explains why those remedies
underperform and which ones do not.

\item \textbf{Transferable guidance} (Sec.~\ref{sec:guidance}), including a
pre-deployment checklist and the observation, measured, that a stabilisation
effective in one regime can be actively harmful in another.
\end{enumerate}

Throughout we report negative and null results at the same level of detail
as positive ones, and we mark clearly which claims are measured, which are
modelled, and which remain open. Section~\ref{sec:setting} describes the
testbed and the surrogate; Section~\ref{sec:protocol} the evaluation
protocol; Sections~\ref{sec:budget} to \ref{sec:mechanism} the results;
Section~\ref{sec:guidance} the discussion and guidance; and
Section~\ref{sec:conclusion} concludes.

\section{Problem setting}\label{sec:setting}

\subsection{The target block}

The host code solves the Einstein equations in the Baumgarte-Shapiro-Shibata-Nakamura (BSSN)
formulation
\citep{shibata1995, baumgarte1999} together with a two-moment (M1)
neutrino radiation-hydrodynamics scheme \citep{shibata2011}; it is the
general-relativistic radiation-hydrodynamics code developed by Kuroda and
collaborators \citep{kuroda2016, kuroda2021, kurodashibata2024,
kurodakawaguchi2025}. The details of the spacetime evolution are not
essential to what follows; what matters computationally is the structure of
the block we target.

For every grid cell and every time step, an inner Newton iteration
(\texttt{NeuMatIntImplicit}) solves the stiff coupling between the local
matter state and the multi-group M1 radiation field. Each call takes as
input the pre-coupling state, namely rest-mass density $\rho$, entropy $s$
(recorded as the equivalent temperature $T$, which the implicit solve does
not search independently), electron fraction $Y_e$, time step $\Delta t$,
conserved energy $\tau$,
momentum $m_1$, and the incoming radiation moments $\mathrm{QSN}_{\rm O}$
(20 energy groups $\times$ 4 four-momentum components $\times$ 3 species),
and returns the post-coupling state ($\mathrm{QSN}_{\rm N}$, $Y_e'$,
$\tau'$, $m_1'$). Each logged record is 493 double-precision values; a
495-value variant additionally logs entropy before and after.

This is a per-cell, state-in/state-out map with no explicit spatial
coupling in its interface, which is precisely the interface shape that makes
a learned surrogate structurally straightforward: no mesh topology, no
neighbour stencil, no time history. The map is nonetheless stiff, and the
radiation update accounts for roughly $80\%$ of the per-call solver time
against $20\%$ for the matter update, measured pre-bounce over some
$21{,}000$ calls and consistent across MPI ranks to within $5\%$.

Two conventions apply throughout. We work per grid cell in cgs units, and we
denote by $t_{\rm pb}$ the time after core bounce, the moment at which the
collapsing core reaches nuclear saturation density and rebounds. The
collapse divides into three regimes with materially different numerical
character (infall, bounce, post-bounce), and we report separately for each
where the distinction matters, since one of our findings is that behaviour
does not transfer between them.

\subsection{Data and surrogate}

We use a single progenitor (a $9.6\,M_\odot$ model, \texttt{z9.6}) with the
DD2 nuclear equation of state \citep{typel2010, hempel2010} and log every
converged call to the coupling solver. The training set (nested-box
refinement level $L_{\rm max}\le7$) contains 2.74M records spanning collapse
through core bounce. As an independent test we generate a second run at full
resolution ($L_{\rm max}$ up to 11, 4.3M records); because it explores the
state space along a different numerical trajectory it is a genuine
out-of-distribution probe rather than a random split.

The surrogate is a residual multilayer perceptron. Inputs are transformed by
$\log_{10}$ ($\rho,T,\Delta t$), a signed logarithm for the
wide-dynamic-range radiation and matter momenta, and per-feature
$z$-scoring; the network predicts residuals in the same transformed space,
with the electron fraction predicted directly. An architecture sweep over
nine configurations selects a width-512, 3-block network ($1.8$M
parameters). Width helps more than depth (width-512/3 beats width-256/6),
LayerNorm improves parameter efficiency, and capacity saturates: a
6.3M-parameter network is \emph{worse} than the 1.8M one, indicating a data-
rather than capacity-limited regime. GELU and SiLU are indistinguishable.
Training uses Adam (initial learning rate $10^{-3}$, reduced on validation
plateau) with a batch size of $8192$ and crash-safe checkpointing, on a
random $90/10$ train/validation split taken at the level of contiguous
data windows (not individual records, to avoid near-duplicate leakage
across the split; Sec.~\ref{sec:offline} discusses a further,
independent-trajectory generalisation test beyond this split). This recipe
is held fixed across every architecture and every retraining in this
paper; only the width, depth, and (in Sec.~\ref{sec:distill}) the random
seed are varied.

For deployment the forward pass was re-implemented in Fortran and verified
against the PyTorch reference to a maximum absolute deviation of
$5.6\times10^{-6}$ and a median relative deviation of $1.7\times10^{-5}$
over $50$ logged records (the maximum relative deviation is large only on
outputs near zero). A later batched implementation reproduced the
scalar path's printed diagnostics exactly over a $65$-step single-threaded
trajectory, its per-call difference lying below print resolution. Agreement at that level is not identity, and
Sec.~\ref{sec:purerep} shows that it does not imply identical trajectories
in general.
We stress this verification because several of our measurements are
differences between configurations, and an unverified inference path would
make those differences uninterpretable.

\section{Evaluation protocol}\label{sec:protocol}

\subsection{Closed-loop substitution}

The decisive evaluation replaces the \texttt{call NeuMatIntImplicit} in the
hydrodynamics sweep with a call to the network and runs a fresh collapse.
The network's outputs set the conserved variables and electron fraction; the
solver's equation-of-state epilogue then recovers the primitive state. We
measure how many steps the simulation survives before a genuine crash, and
whether the state remains physical.

Two properties of this protocol matter for interpreting what follows. It is
a substitution test, not a coupled-training test: the surrogate is trained
offline and deployed frozen. And the failure criterion is the host code's
own guards (equation-of-state range checks, a causality check on
characteristic wave speeds), so a ``crash'' is a physically meaningful
violation rather than a numerical convenience.

\subsection{Identity ablations}

To localise failures we selectively replace individual network outputs with
the identity map, so that the coupling does nothing for that channel while
the network continues to drive the others. This isolates which predicted
channel is responsible for a given failure, and it provides a control: with
every channel set to identity the network is inert, and the framework itself
can be validated independently of the surrogate.

\subsection{Timing methodology}

End-to-end comparisons are matched-window: identical starting state,
identical number of steps, matched thread and rank counts, with startup
excluded. Where node contention could confound a comparison we run the pair
in both orders and report agreement. Profiling uses exclusive (self-time)
timers, so that a routine is charged only for time spent outside its
instrumented callees; we validate the decomposition by checking that
disjoint leaves sum to the measured sweep duration within $0.2\%$. This
validation is not incidental. Nested timers double-count, and a
decomposition that does not sum to the whole cannot support an Amdahl
argument.

\subsection{Seed protocol}

Closed-loop survival is a chaotic quantity, and single runs are not reliable
estimates of it. Where we compare configurations we retrain each five or
more times, varying only the random seed (weight initialisation and shuffle
order) with an otherwise identical recipe and an identical data partition,
and restart all runs from the same verified state. We report distributions
and rank tests rather than point comparisons.

One protocol detail is specific to this system and worth stating because it
determines what a ``seed'' can mean. The simulation itself is fully
deterministic: it contains no random number generator, and five identical
repeat runs of the same model produce identical step counts. Seed variation
therefore varies the \emph{training} initialisation, not the trajectory, and
the resulting spread measures sensitivity of the closed loop to which model
was learned, not to numerical noise in the host code.

\section{The acceleration budget}\label{sec:budget}

We begin with the quantity that bounds everything else, because it reframes
what a surrogate for this block can deliver before any question of accuracy
or stability arises.

\subsection{Per-call cost is not a runtime share}

The target routine is the most expensive physics routine per call in this
code. It is natural, and it was natural to us, to read that as implying it
dominates runtime. It does not, and the two quantities are related by the
call count and by everything else the time step does.

We record this explicitly because the conflation had consequences in our own
work. An early cost figure of $3.4\times10^{-4}\,$s per call was carried
forward in planning as though it were a fraction of total runtime, which
supported an expected ceiling of roughly $2.7\times$. That expectation
survived until an end-to-end measurement contradicted it, at which point the
profiling below became the critical path rather than a supporting
measurement. The distinction is elementary; the failure mode is that a
per-call benchmark is easy to obtain and a validated runtime share is not,
so the former tends to stand in for the latter.

\subsection{Exclusive self-time profile}

Table~\ref{tab:profile} reports the exclusive self-time decomposition of one
time step for the configuration used throughout this work, both pre-bounce
and restarted at $t_{\rm pb}=20\,$ms. Six categories are large enough to
discuss individually; the remainder, $6$--$7\%$ in every phase, is made up
of the conformal-factor rescaling and the conserved-to-primitive recovery,
which are of comparable size and together account for most of it, together
with the spacetime evolution, the gravitational source term and state
storage. None of it is idle time, and none of it changes which routine
dominates.

\begin{table}[t]
\centering
\caption{Exclusive self-time breakdown of one time step, as a percentage of
mean wall clock across MPI ranks, for the three phases of the collapse.
Leaves are disjoint; the six named categories are the ones large enough to
discuss individually, and the residual row recovers the remainder so that
each column sums to $100\%$. The lower block gives the solver's
share in two other framings and the mean number of Newton iterations per
cell, which is the physical reason the solver cost varies across phases.}
\label{tab:profile}
\begin{tabular}{lrrr}
\toprule
Routine (disjoint leaf) & Pre-bounce & Bounce & Post-bounce \\
\midrule
Ghost exchange, mostly idle wait & $30.8$ & $27.8$ & $34.6$ \\
PPM advection                    & $20.3$ & $27.6$ & $21.7$ \\
Ghost exchange, genuine MPI      & $13.5$ & $9.1$  & $12.0$ \\
Loop overhead residual           & $11.0$ & $10.1$ & $11.8$ \\
\textbf{Implicit coupling solve} & $\mathbf{7.5}$ & $\mathbf{16.1}$ & $\mathbf{10.5}$ \\
Periodic I/O                     & $9.8$  & $2.5$  & $3.1$ \\
Other genuine work \& untabulated & $7.1$ & $6.8$  & $6.3$ \\
\midrule
Solver, \% of critical-rank wall & $11.9$ & $17.7$ & $16.9$ \\
Solver, \% of pure compute       & $21.5$ & $31.9$ & $27.4$ \\
Newton iterations per cell, mean & $0.37$ & $2.29$ & $1.56$ \\
Newton iterations per cell, max  & $1$    & $30$   & $12$ \\
\bottomrule
\end{tabular}
\end{table}

The coupling solve is $10.5\%$ of mean wall clock post-bounce. Because step
duration is set by the slowest rank, the figure relevant to Amdahl's law is
its share on that critical rank, $16.9\%$. The mean is depressed by idle
wait, which is not work: the same-level ghost-exchange timer records $52\,$s
on the dense-core ranks, which hold the fewest cells and therefore finish
early and wait, against $0.44\,$s on the outer-zone rank that holds seven
times as many cells and sets the pace, so it measures ranks waiting rather
than messages in flight. That
$0.44\,$s rank is the critical one, so the $16.9\%$ figure is not depressed
by idle wait at all; it is depressed by genuine communication, periodic I/O
and loop overhead. Charging the solver against pure compute alone, the
fraction an idealised perfectly balanced single-rank code would exhibit,
raises it to $27.4\%$ post-bounce and $21.5\%$ pre-bounce. That denominator
excludes the ghost-exchange leaf entirely; post-bounce $22.7\%$ of that leaf
is not wait but ghost-cell work, chiefly equation-of-state evaluations and
Cartoon interpolation, a technique that imposes axial or spherical symmetry
in a Cartesian calculation by reconstructing off-plane values through rotation
and interpolation. Counting it lowers the solver's pure-compute share
to $22.7\%$ and the idealised ceiling from $1.38\times$ to $1.29\times$. We
quote the more favourable convention throughout.

The share varies across phases, and it peaks where physics predicts. At
bounce the coupling is stiffest and the Newton iteration needs a mean of
$2.29$ iterations per cell against $1.56$ post-bounce and $0.37$ during
infall, with maxima of $30$, $12$ and $1$. Wall-clock share tracks this
closely, reaching $16.1\%$ of mean wall clock. On the critical rank the
share is $17.7\%$, only marginally above the post-bounce $16.9\%$: the rank
carrying the most solver work at bounce is not the rank that sets the step
duration, so the solver's prominence there does not translate into a higher
ceiling. Bounce is thus the most favourable phase for a surrogate anywhere
in this problem, but only just, with a ceiling of $1.21\times$ against
$1.20\times$ post-bounce. Nor does it change the conclusion for a further
reason: the bounce window spans about $3700$ of the $53{,}203$ steps to
bounce, some $7\%$ of the collapse.

\subsection{The resulting bound, and its confirmation}

A free surrogate, of perfect accuracy and unconditional stability, replacing
this routine entirely, is bounded by $1/(1-f)$: about $1.20\times$ at
$f=0.169$ in the configuration studied here, and about $1.38\times$ at
$f=0.274$ in the balanced idealisation.

This bound binds in practice. The batched post-bounce surrogate, a
width-$256$ network, costs $5.9\times10^{-5}\,$s per cell
against the solver's $3.4\times10^{-4}\,$s, a factor $5.8$ cheaper per call,
and merely \emph{ties} the solver end-to-end ($0.296$ against $0.318\,$s per
step, matched window, startup excluded).

We checked the obvious objection, that the low fraction is an artefact of an
over-decomposed benchmark in which surface-scaling communication swamps
volume-scaling physics. Growing the per-rank volume eightfold moves the
solver share only from $7.5\%$ to $11.3\%$, while the genuinely
surface-bound communication term falls from $13.5\%$ to $8.4\%$: a real but
minority effect with no path towards a solver-dominated regime. We state the
fraction as a property of this code and configuration and do not extrapolate
it to production scale, but nothing in the volume scaling suggests the
routine approaches majority cost there either.

\subsection{The dominant cost is not the solver}

The coupling solve is not the largest lever on wall clock in this code. Load imbalance is. Charging each rank only for
genuine work (solver, advection, conserved-to-primitive recovery,
gravitational source), the busiest rank does $1.82\times$ the mean rank's
compute post-bounce ($53.2$ against $29.3\,$s), and idle time tracks compute
inversely. Modelling a perfectly balanced decomposition and bracketing the
residual overhead between optimistic and conservative assumptions yields
$1.13$ to $1.40\times$ post-bounce and $1.11$ to $1.27\times$ pre-bounce.
Combining both levers gives $1.29$ to $1.64\times$, slightly
super-multiplicative because balancing shrinks the compute pool against
which the solver's share is charged.

Two qualifications keep this from being oversold. The balance estimate is a
\emph{model} with stated bracketing assumptions, whereas the surrogate
number is measured; at the conservative end the two levers are nearly equal.
More instructively, removing the solver from the current bottleneck rank
does not simply remove its cost from the step: the critical path migrates to
a different, idle-dominated rank. The surrogate's realised gain is therefore
smaller than a naive application of Amdahl's law to the aggregate share
predicts, which is exactly what the measured tie shows.

Whether the imbalance can be removed is a separate question with a
discouraging answer for a quick fix. The decomposition is static and
contiguous in radius, one block per rank, assigned once; the rank count is
pinned to a compile-time constant and the ghost-exchange routines fail below
four ranks. A non-uniform static block width chosen from the density profile
is architecturally compatible but would require promoting a global
grid-spacing scalar to a per-block quantity and auditing every routine
indexed off it. We note it because it is the dominant lever and because it
is not a machine-learning problem at all.

\subsection{Where learned surrogates have delivered savings}

The literature pattern is instructive and consistent with the bound above.
The clearest success replaces a component that \emph{dominates} runtime.
\citet{fan2022maestroex} substitute a network for the stiff nuclear reaction
integrator in a low-Mach stellar hydrodynamics code where reactions occupy
$59.5\%$ of total runtime; the surrogate reduces that share to $29.4\%$ and
speeds the reaction kernel by $3.14$ to $3.45$ times. Even there the
reported figure is a kernel speedup, with no end-to-end wall clock for the
full simulation. The other success mode replaces the entire time step, as in
learned weather emulators, where no Amdahl ceiling applies because nothing
of the original solver remains.

The case closest to ours points the same way. \citet{dieselhorst2021c2p}
learn the conservative-to-primitive inversion in relativistic hydrodynamics,
a minority-cost inner root find much like our coupling solve, and report
that the network accelerates variable recovery ``by more than an order of
magnitude''. That is a per-call inference benchmark; the paper gives neither
the inversion's share of runtime nor a measured wall clock for a running
simulation. \citet{zhang2025deepode} are explicit that their reported factor
is a conservative per-call estimate and that a realistic acceleration
baseline remains to be established. We could find no published case
in which replacing a minority-cost inner block produced a measured
end-to-end speedup. Our numbers explain why that absence is unsurprising
rather than accidental.

Two further directions differ in kind rather than degree. First, several
groups accelerate the exact solve rather than replacing it.
\citet{laiu2020thornado} replace dense-Jacobian Newton iteration for the
same neutrino-matter coupling step with a Jacobian-free, GPU-batched,
Anderson-accelerated fixed-point solve, reporting up to $100\times$ against
a single CPU core. This requires restructuring the solve to operate across
the whole grid at once rather than cell by cell, and uses no learned
component; it is the clearest evidence we found that an order-of-magnitude
gain is available in this exact block, just not through a drop-in surrogate.
That line has since been carried into a production framework
\citep{endeve2026thornado}, again without a learned component.
Learned preconditioners for the underlying linear solves
\citep{li2023precond, trifonov2024precond} are a related non-replacement
direction, though neither reports a figure directly comparable to ours.
Second, hybrid schemes that correct a coarsened solver with a learned term
\citep{kochkov2021mlcfd, um2020solverloop} can yield large measured speedups:
\citet{kochkov2021mlcfd} report up to $80\times$, matching an $8$ to
$10\times$ finer classical solve, but against a coarser-resolution
classical baseline rather than a same-fidelity one. That is accuracy per
unit compute at fixed cost, not the same-fidelity replacement tested here,
so the results are not in tension. Physics-informed and neural-ODE
approaches to stiff microphysics \citep{raissi2019pinn} embed the governing
equations in the loss rather than fitting solver input-output pairs; we are
not aware of an application to the neutrino-matter coupling specifically, and
note it as an architecturally distinct alternative rather than a comparable
result.

Machine learning has been applied to neutrino transport itself, but at a
different point in the algorithm than the one we target, which is why those
results neither support nor contradict ours.
\citet{harada2022eddington} and \citet{takahashi2025lightgbm} learn the
Eddington tensor, that is the \emph{closure} that the two-moment scheme
requires, replacing an analytic approximation by a fitted one; the aim is
fidelity to a Boltzmann reference rather than wall clock, and the learned
closure enters the same implicit solve we time here rather than replacing
it. \citet{abbar2025mlffi} and \citet{richers2024asymptotic} use learned
models for fast flavour instabilities, a physical process our code does not
resolve at all; notably \citet{richers2024asymptotic} find that an analytic
subgrid model generalises better than the learned one, which is the same
ordering of learned against constructed methods that our gate analysis
produces by a different route. We are not aware of published work that
replaces the neutrino-matter coupling solve itself with a surrogate inside a
running simulation, which is the configuration measured here.

The reporting pattern behind this literature has itself been examined.
\citet{mcgreivy2024weak} survey machine-learning solvers for fluid-related
partial differential equations and find that $79\%$ of articles claiming to
outperform a standard numerical method compare against a weak baseline, with
outcome- and publication-reporting biases suppressing negative results. Our
protocol is a direct response to that finding: the baseline is the
production solver in its own code at the same fidelity, the comparison is
end-to-end rather than per call, and the result is negative.

\subsection{On hardware}\label{sec:gpu}

All measurements here are CPU-only. We have no GPU infrastructure for this
code, and a port of the Fortran solver is well beyond this work's scope. As
an analytical estimate rather than a measurement, it is worth asking whether
a GPU changes the conclusions.

A partial port of only the coupling block, with the rest of the step left on
the CPU, would require transferring per-cell state across the host-device
boundary every step; at this routine's small per-cell payload, transfer and
kernel-launch overhead could plausibly erase any gain unless the whole
per-rank cell array is batched in one call, which is precisely the
restructuring \citet{laiu2020thornado} perform for the exact solver. Using
the break-even formula of Sec.~\ref{sec:gate} to bound the best case for the
gate, driving its cost to zero ($g\to0$, an optimistic proxy for a
well-batched GPU implementation of a $249$-dimensional quadratic form) while
holding the best measured deferral ($d=0.981$) and network cost ratio
($r=0.17$) fixed gives $S\approx1.002$ pre-bounce and $S\approx1.003$
post-bounce. That is a swing from the measured $0.94$ to $0.96\times$
slowdown to a two-to-three-per-mille gain, not a qualitative change.

The reason is structural rather than computational: the deferral fraction,
not the gate's cost, is the binding constraint, and deferral is set by how
far the closed loop has drifted off the training manifold, a property of the
physics and the training data that hardware speed does not change. A GPU
port of the exact solver is not subject to this limit and remains the
credible route to a real speedup in this block. If anything, making the
coupling routine cheaper relative to the rest of the step would shrink $f$
further and tighten, not loosen, the Amdahl ceiling on any surrogate for it.

\section{Per-step accuracy is not a deployment metric}\label{sec:offline}

\subsection{High offline accuracy carries no deployment signal}

\begin{figure}[t]\centering
\includegraphics[width=\textwidth]{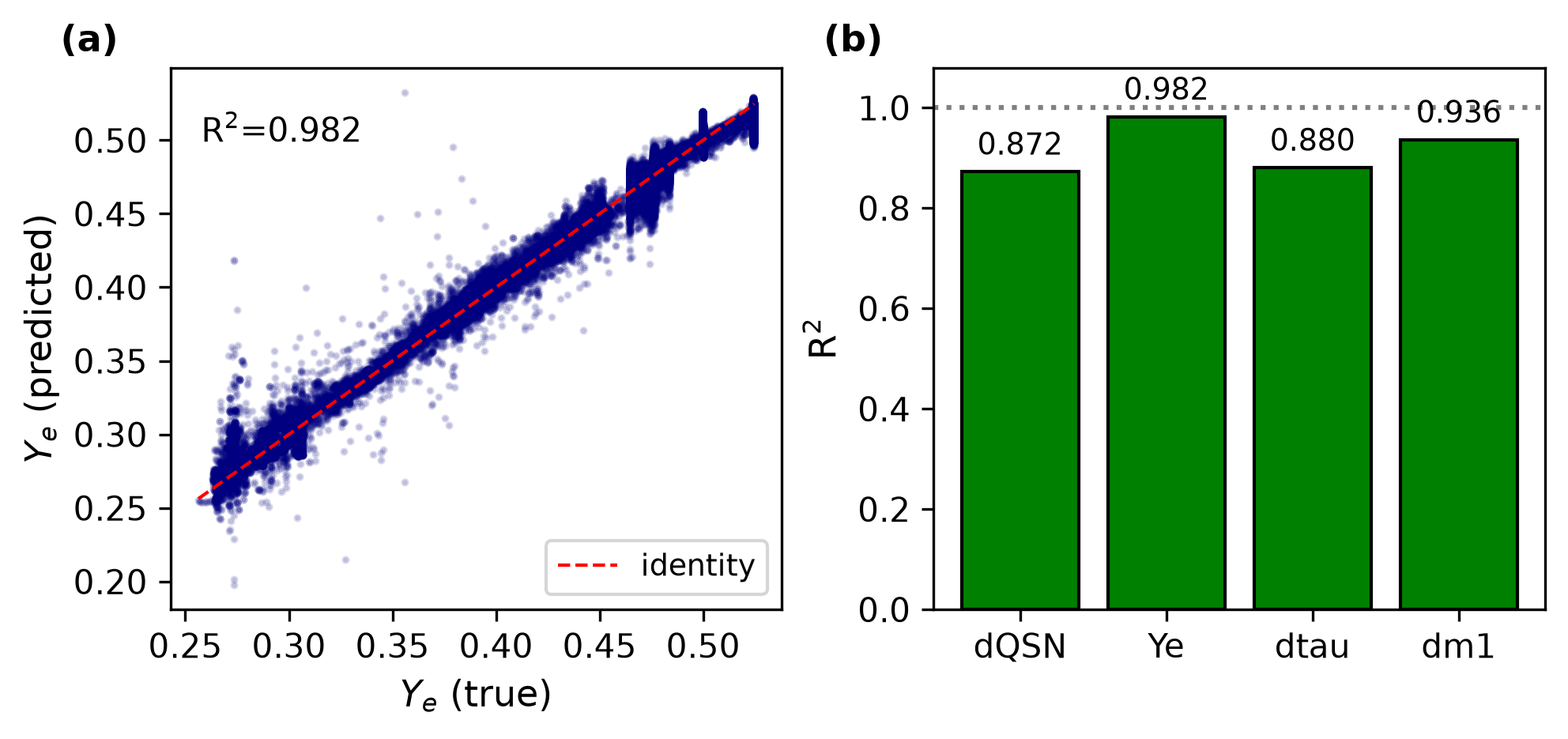}
\caption{Per-step accuracy of the wide model on a held-out split of the
training distribution. \emph{(a)} Predicted against true electron fraction
with the identity line ($R^2=0.982$, mean absolute error
$3.4\times10^{-3}$). \emph{(b)} Held-out $R^2$ per predicted quantity. These in-distribution figures are optimistic, and
the remainder of this section shows they carry no ranking signal for
deployment.}
\label{fig:perf}
\end{figure}

On a held-out split of the training distribution the wide model reaches
$R^2=0.982$ for $Y_e$ and $0.87$ to $0.94$ for the radiation, energy and
momentum changes, with an electron-fraction mean absolute error of
$3.4\times10^{-3}$ (Fig.~\ref{fig:perf}). These numbers are optimistic: the
data are a single highly autocorrelated trajectory, so a random split leaks
near-duplicate states between train and test.

Evaluated on the never-trained full-resolution trajectory, the electron
fraction generalises well ($R^2=0.90$ to $0.98$) but the energy change
collapses ($R^2=0.06$ to $0.32$) and the momentum change degrades
($R^2=0.29$ to $0.45$). The network has memorised the energy coupling of the
training trajectory rather than learning a transferable map. The natural
hypothesis, insufficient data diversity, we tested directly by training on
two resolutions jointly and evaluating on a held-out dense regime: the
energy $R^2$ did \emph{not} improve; it worsened ($-0.08\to-0.92$). More
data of the same kind does not help.

In the live simulation the naive surrogate crashes within tens of steps. A
network with $R^2=0.98$ per call terminates the simulation within two steps.

\subsection{Seed variance forbids single-run comparison}

Before comparing configurations we must establish what a difference means.
We retrained two configurations five times each, varying only the seed. A
narrow surrogate trained on the two innermost, densest cells survives $70$,
$73$, $75$, $89$ and $102$ steps (median $75$, interquartile range $16$). A
broad surrogate trained on $128$ cells spanning the whole radial domain
survives $13$, $21$, $21$, $26$ and $29$ steps (median $21$, interquartile
range $5$). The distributions do not overlap: the worst specialist seed
exceeds the best broad seed. A Mann-Whitney rank test gives complete
separation ($U=25$ of $25$, exact two-sided $p=0.008$); with five draws per
group this quantifies only that no overlap was observed.

The spread is substantial in absolute terms, and single-run survival counts
should not be compared when they differ by less than it. This discipline
immediately overturns two of our own earlier conclusions: doubling network
width from $256$ to $512$, a $3.5\times$ increase in parameters, and
reweighting the loss so that crash-critical core cells
receive half the gradient mass, each looked decisively harmful on a single
seed. Repeated over five seeds, neither differs from the unmodified broad
model (medians $22$ and $22$ against $21$; $p=0.84$ and $p=0.52$). Their
apparent harm was seed noise, and we report them as null results.

One intervention survives: routing each cell by local density to one of two
narrow experts, trained respectively on the dense core and the outer zone.
This raises the median to $47$ steps, significantly above the broad model
($p=0.016$) at no additional inference cost, since exactly one expert is
evaluated per cell. It remains cleanly separated from the single narrow
specialist ($p=0.008$), so specialisation, not routing, carries most of the
benefit.

Two features of these runs are entirely seed-independent and carry the
mechanism. The crash location: the narrow specialist fails in the outer,
low-density region it never saw during training, in all five seeds, while
the broad model fails at the dense core, in all five. Breadth of training
domain does not buy competence at the core; it costs it. And the physical
health of the terminal state: only three of five specialist seeds carry
central density monotonically upward to the causality abort, while
\emph{none} of the five broad seeds does, every one reversing density before
failing. Judged by physical validity rather than step count, the broad
configuration never once reaches a healthy state.

A caution follows, and it bears on scoring generally. Ranking five
configurations by median survival and by fraction of seeds reaching a
physically healthy terminal state gives \emph{different} orderings: the
wider model ranks near the bottom by step count yet matches the
specialist for clean terminal physics, while the routed model ranks second
by step count and near the bottom by physical health. Step count and
physical validity are only partly coupled, and a surrogate evaluated on the
former alone can look better than it is.

\subsection{Offline error carries no usable ranking signal}

Scored on a single common held-out set spanning both dense core and outer
zone, verified by content fingerprinting to be excluded from every training
set, the narrow model has the \emph{worst} one-step electron-fraction error,
$0.150$, while the broad model has the best, $0.023$; their survival medians
are $75$ and $21$. Restricting the error to the dense core, where the broad
model actually fails, does not change this.

\begin{figure}[t]\centering
\includegraphics[width=\textwidth]{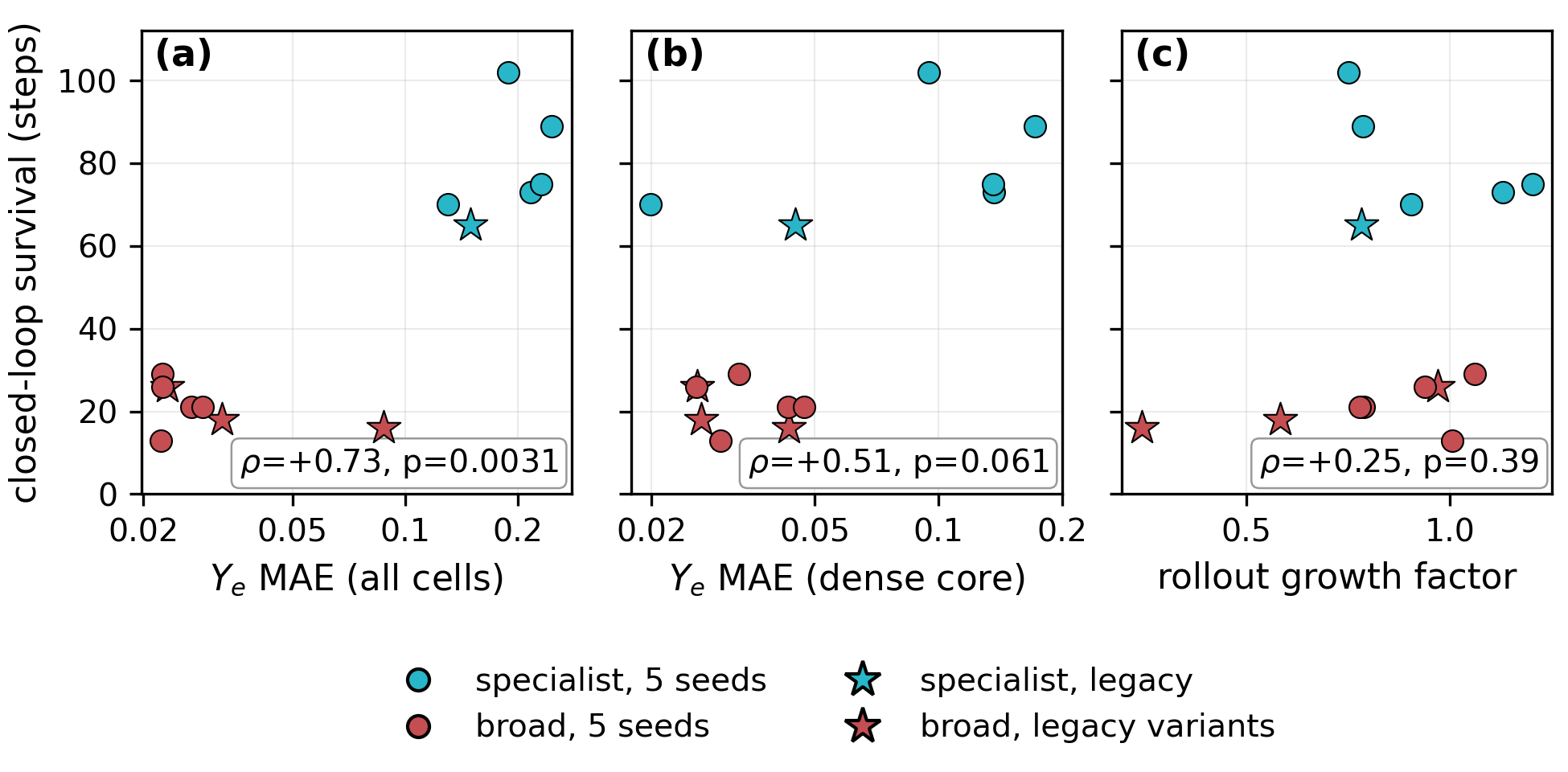}
\caption{Offline error does not rank surrogates by closed-loop survival.
Fourteen independently trained networks, all scored on one common held-out
set (verified excluded from every training set), against closed-loop steps
survived. The points separate into two clusters, narrow specialists (cyan,
upper right) and broad models (red, lower left), and the apparently strong
pooled correlation in the left panel is produced by that separation: within
either cluster there is no trend, and the partial rank correlation
controlling for cluster membership is $\rho=-0.04$ ($p=0.89$). Restricting
the error to the dense core (centre), where the broad models actually fail,
does not help, nor does the multi-step rollout error growth factor (right).
Circles are the five seeds of each configuration; stars are earlier
single-seed models.}
\label{fig:offlinesurvival}
\end{figure}

Pooling fourteen independently trained surrogates scored on that common set,
the Spearman rank correlation coefficient between one-step error and
survival is $\rho=+0.73$
($p=0.003$, $n=14$), significant and of the sign opposite to the one a
practitioner would assume. It would be tempting to report this as evidence
that offline error is actively anti-predictive. It is not, and the
distinction matters. The fourteen models fall into two families, narrow and
broad, and the families differ in both quantities at once: the narrow family
has higher offline error \emph{and} longer survival. Controlling for family,
the partial rank correlation collapses to $\rho=-0.04$ ($p=0.89$;
Fig.~\ref{fig:offlinesurvival}). Within
families it is not statistically significant and changes sign ($\rho=+0.60$, $p=0.21$
narrow; $\rho=-0.31$, $p=0.45$ broad). The pooled coefficient restates the
difference between families rather than measuring any property of the
metric. The multi-step rollout error growth factor is uninformative
throughout ($\rho=+0.25$, $p=0.39$ pooled), as is any of these metrics for
predicting whether the terminal state is physically healthy (point-biserial
$r=0.52$, $p=0.06$, at best borderline).

A matched pair makes the resolution question concrete. Two checkpoints of
the same network, trained from the same seed on the same data with the same
recipe, differ in their weights (a consequence of an irreproducible
divergence during training) and survive $56$ and $397$ closed-loop steps, a
factor of seven. Scored on the common set, their one-step electron-fraction
errors differ by $2.95\%$. A second pair differs by $0.85\%$ offline while
surviving $326$ and $389$ steps. In both pairs the longer-surviving model
does have the slightly lower offline error, so the metric is not pointing
the wrong way; it simply has no resolution. Three per cent of offline signal
stands against a sevenfold difference in the quantity that matters. We
report this as illustration rather than evidence in its own right: two pairs
are an anecdote, and they arise from a training divergence we could not
deliberately reproduce.

The cross-model comparison only became meaningful once evaluation sets were
unified. Judged on their own evaluation sets, these models appear to show a
clean monotone relationship between one-step error and survival, an artefact
of the narrow-domain model being scored on the two easy cells it was trained
on. The practical consequence is therefore sharper than a simple inversion:
offline error carries \emph{no usable ranking signal within a
configuration}, and the only pooled correlation it exhibits is a confound.
Selection must be performed in the loop, on distributions rather than single
runs.

\section{Closed-loop failure: diagnosis}\label{sec:diagnosis}

\subsection{Two obstacles, separated by ablation}

Identity ablations localise the failure precisely (Table~\ref{tab:ablation}).

\begin{table}[t]
\centering
\caption{Identity ablations isolate two obstacles. The framework itself is
sound ($300$ clean steps with an inert network). Predicting the conserved
energy and momentum causes an acute density collapse at step $2$; with those
held fixed, a slow electron-fraction drift terminates the run near step
$170$.}
\label{tab:ablation}
\footnotesize
\setlength{\tabcolsep}{4pt}
\begin{tabular}{lcc}
\toprule
Configuration & Steps survived & Density \\
\midrule
Full identity (network inert) & 300 (clean) & stable \\
Network drives $Y_e$+QSN+entropy only & $\sim\!170$ & stable, then $Y_e$ drift \\
Network drives everything (including $\tau,m_1$) & 2 & collapse $100\times$ \\
\bottomrule
\end{tabular}
\end{table}

\subsection{The acute obstacle: an unlearnable target}

\begin{figure}[t]\centering
\includegraphics[width=0.72\textwidth]{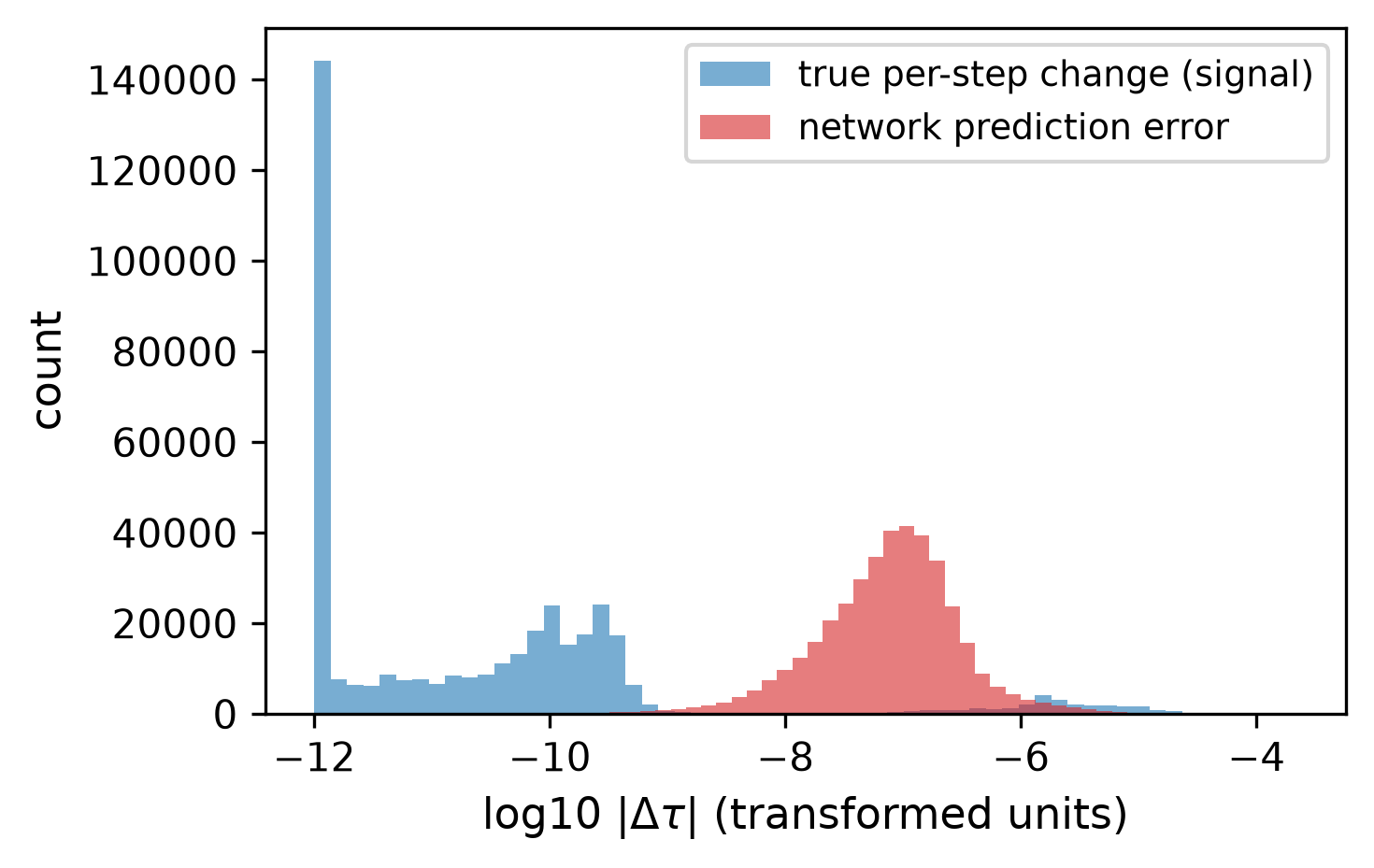}
\caption{Signal-to-noise of the energy channel. The true per-step energy
change is $\sim\!5500\times$ smaller than the network error: the increment
is below the learnable resolution.}
\label{fig:snr}
\end{figure}

The acute collapse has a quantitative explanation, and it is a general
lesson about which quantities should be regressed at all. The per-step
change of the conserved matter energy is not merely small but, for most
steps, \emph{unrepresentable}: across $10^{5}$ logged cell updates the
relative change $|\Delta\tau/\tau|$ is bit-exact zero in $59\%$ of steps,
its median is exactly zero, its $99$th percentile is $5.7\times10^{-5}$, and
it never exceeds $3\times10^{-4}$. In the network's transformed target units
the median signal is $1.5\times10^{-11}$ against a prediction error of
$8.3\times10^{-8}$, a factor $\sim\!5500$ (Fig.~\ref{fig:snr}).

A regression network cannot represent a target that is exactly zero most of
the time and minute otherwise; it emits small non-zero noise everywhere. The
downstream conserved-to-primitive inversion then amplifies this spurious
increment: a $5\%$ energy error drives a $\sim\!100\times$ density collapse
within one to two steps, because the coupled recovery ties density, energy
and Lorentz factor together.

Seen through a conservation lens the same mismatch appears as a
$\sim\!10^{7}$ energy-conservation residual relative to the solver, and it
would be easy to report that as a physics violation requiring a
conservation-enforcing architecture. It is not: it is the same
sub-resolution mismatch, a representable radiation-energy change set against
a matter-energy change below floating-point resolution. A
conservation-by-construction layer is therefore structurally infeasible
here, and the correct remedy is different in kind.

\subsection{The remedy: predict the learnable state, derive the rest}

Because the conserved energy increment is unlearnable but the thermodynamic
state is not, we form the energy through the equation of state rather than
predicting it. The network outputs $Y_e$ and entropy; the conserved energy
is then computed exactly as the solver does, from
$(\rho,Y_e',S')$, consistent with the conserved rest-mass density by
construction. This removes the density collapse and extends closed-loop
survival from $2$ to $102$ steps with a stable density profile.

This is the one obstacle we solve outright, and it generalises: when a
target's true increment lies below the achievable regression error, do not
regress it. Predict the quantities that carry signal and recover the rest
through the exact algebraic relations the solver already implements, trading
a noisy regression for a consistent map.

\subsection{The chronic obstacle: covariate shift}

\begin{figure}[t]\centering
\includegraphics[width=\textwidth]{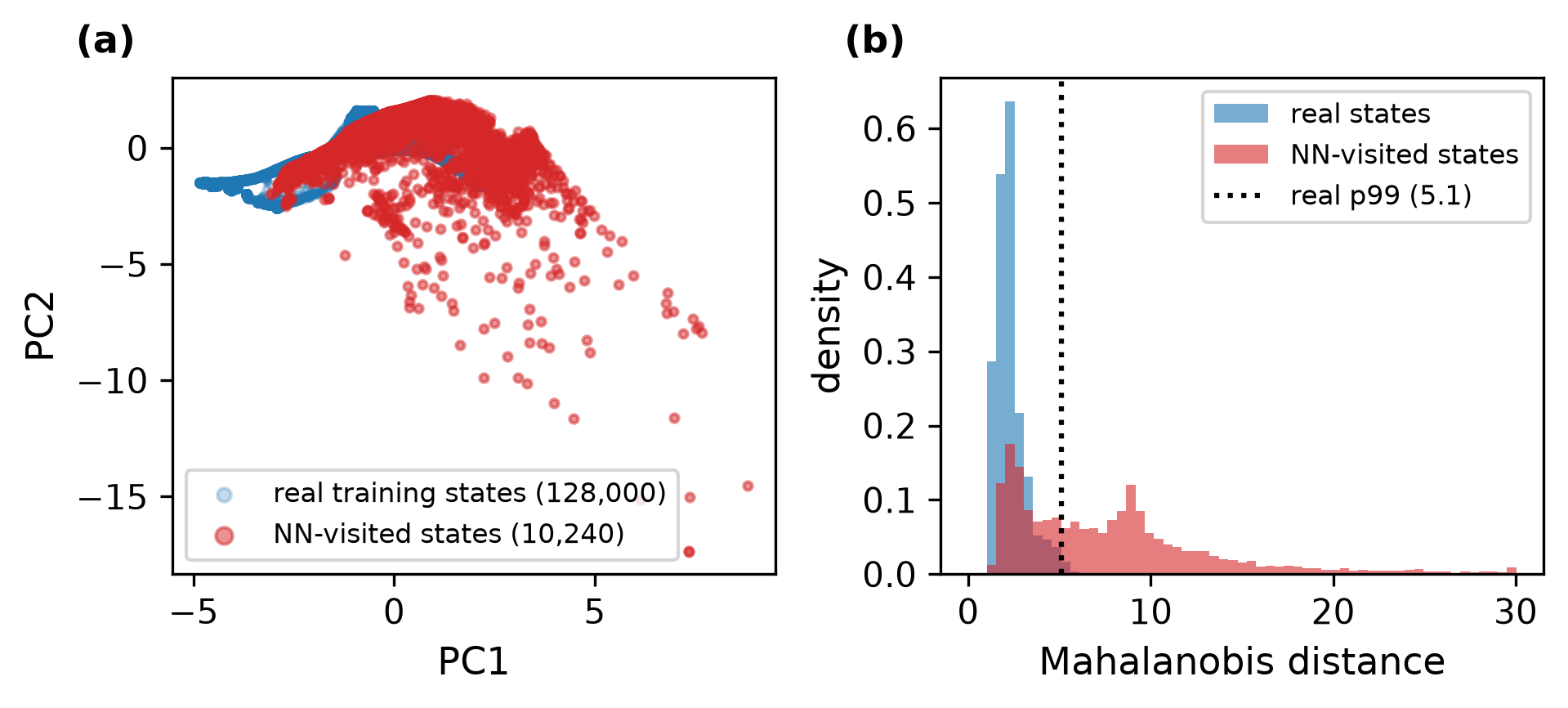}
\caption{The closed-loop drift is covariate shift. Network-visited states
(red) lie far off the real-data manifold (blue): $73\times$ the real-to-real
nearest-neighbour distance, $61\%$ beyond the $99$th-percentile Mahalanobis
shell, with unphysical entropies ($\le62$ versus a real maximum of $7.8$)
that no real trajectory occupies.}
\label{fig:covshift}
\end{figure}

With the acute collapse removed, the surviving failure is a slow
electron-fraction drift. The states the network-driven loop visits are
massively off the training manifold: their $5$-nearest-neighbour distance to
the real-data manifold is $73\times$ larger than real-to-real, $61\%$ lie
beyond the $99$th-percentile Mahalanobis shell, and their entropy reaches
$62$ against a real maximum of $7.8$, thermodynamic states no real cell
trajectory ever occupies (Fig.~\ref{fig:covshift}).

This explains why more real data cannot help: a model trained on the full
$2.3$M all-cell dataset (density span $5\times10^{4}$, $17\%$ of records in
the crash regime) still drifts and crashes at the same $\sim\!100$ steps.
The visited states lie in no real trajectory, so only collecting them from
the network's own rollout can cover them.

\subsection{The failure is distributed, not localisable}

The crash is not attributable to any single channel. Clamping the electron
fraction, clamping the entropy (the most off-manifold variable, reaching
$\sim\!8\times$ its physical maximum), and flooring the density guard that
formally aborts the run each leave the crash at the same $\sim\!550$ steps:
the abort merely migrates to the next guard. The instability is a
distributed off-manifold drift of the whole state, which is why only a
global correction helps and why channel-wise clamping is not a remedy.

\section{Stabilisation strategies and their limits}\label{sec:stabilise}

We attacked the drift from four directions. None removed it, and each
confirms the same diagnosis from a different angle.

\paragraph{Multi-step rollout training}
Training on an eight-step rollout of a single trajectory did not measurably
improve long-horizon stability: one-step and rollout-trained models diverge
almost identically (mean closed-loop $Y_e$ error $0.61$). A curriculum along
one trajectory teaches the network nothing about the off-trajectory states
the loop actually visits.

\paragraph{Periodic solver re-anchoring}
Inserting one true solver step every $N$ network steps does not cure the
drift. With the best aggregated model (horizon $557$), re-anchoring as often
as every $50$ steps still crashes at the same $\sim\!557$: by the time a
re-anchor fires the drift is spread across the whole state, so a single
coupling correction cannot pull it back. A cadence small enough to matter,
$N\sim3$, would in any case forfeit the speedup.

\begin{figure}[t]\centering
\includegraphics[width=0.7\textwidth]{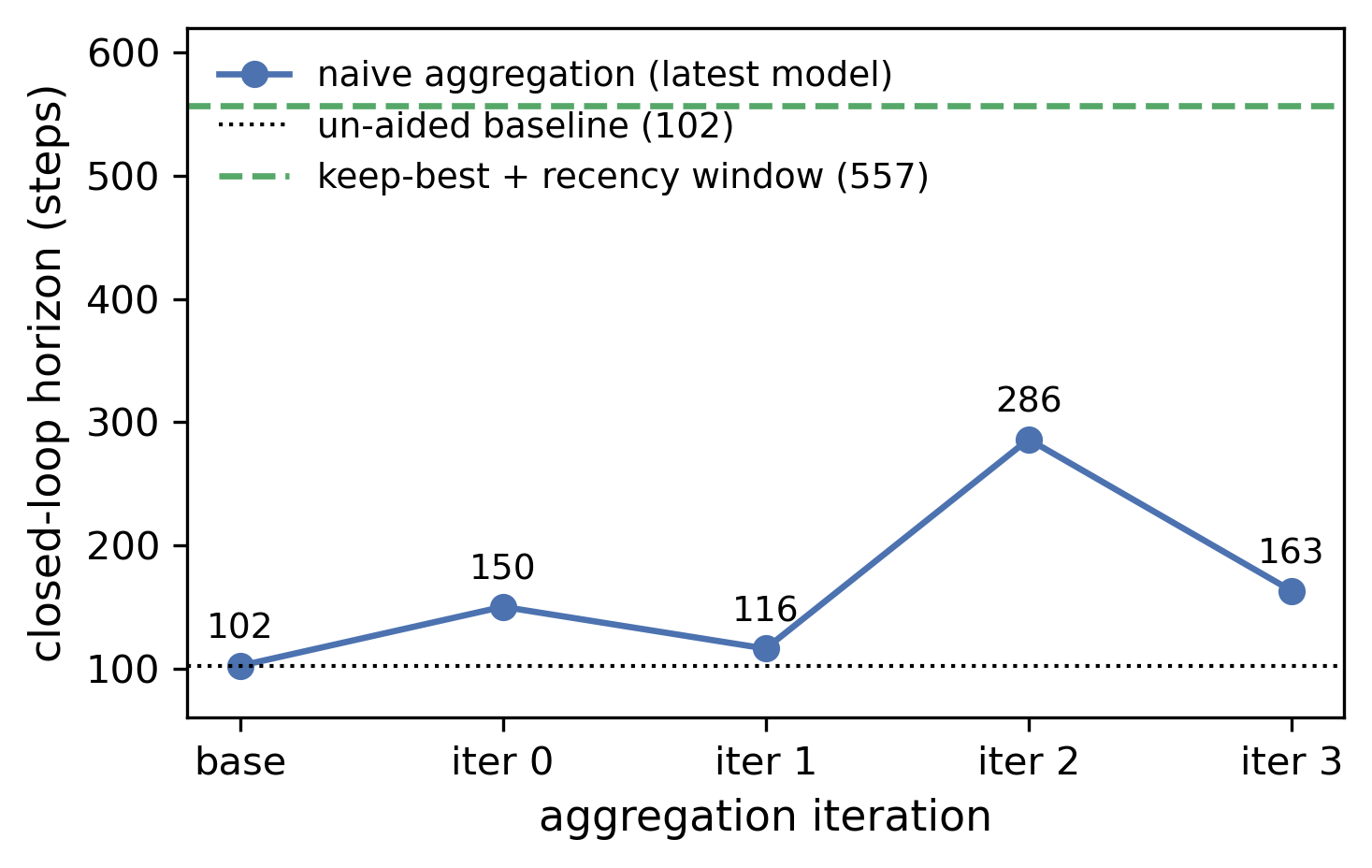}
\caption{Closed-loop horizon (steps to a genuine crash) versus aggregation
iteration. Naive aggregation oscillates without converging (best $286$,
output model worse); keep-best plus a recency window reaches $557$ steps,
$5.5\times$ the baseline ($102$), with $Y_e$ held physical.}
\label{fig:dagger}
\end{figure}

\paragraph{Dataset aggregation}
We instrumented the loop to log the off-trajectory states it visits,
labelled them with the true solver, filtered the non-finite near-crash
cascade ($78$k$\to71$k records; keeping them, or up-weighting them
$4\times$, instead regresses the model to a step-2 crash), and retrained,
verifying deployed weights by checksum at every step. Naive iteration,
retraining on all aggregated states and deploying the latest model, raises
the horizon but does not converge: over four rounds it oscillates
$102\to150\to116\to286\to163$ (all genuine crashes), and the procedure's
output, the \emph{last} model, is not its best. Two corrections fix this:
keep the best-horizon model across rounds, and weight recently visited
states over old ones. The improved scheme reaches $557$ steps ($5.5\times$
the baseline) while holding $Y_e$ physical
(Fig.~\ref{fig:dagger}), following \citet{dagger}. The loop nonetheless
crashes eventually: aggregation substantially \emph{delays} the
covariate-shift divergence without removing it.

\paragraph{Contractivity}
The one model-side lever that helps at no inference cost is contractivity:
penalising the Jacobian norm of the coupling map with respect to the
fed-back variables lengthens the horizon by $\sim\!36\%$ at unchanged
per-step accuracy, and it stacks with aggregation. Stacked, the two reach
$2625$ closed-loop steps as a pure replacement, $26\times$ the un-aided
baseline, with a physical electron fraction throughout. (A
less-characterised run of the same configuration exceeded $4000$
steps.) This is the correct target for a
directed, feedback-amplified error: it constrains how the map responds to
its own recirculated output rather than damping input noise.

\begin{figure}[t]\centering
\includegraphics[width=0.78\textwidth]{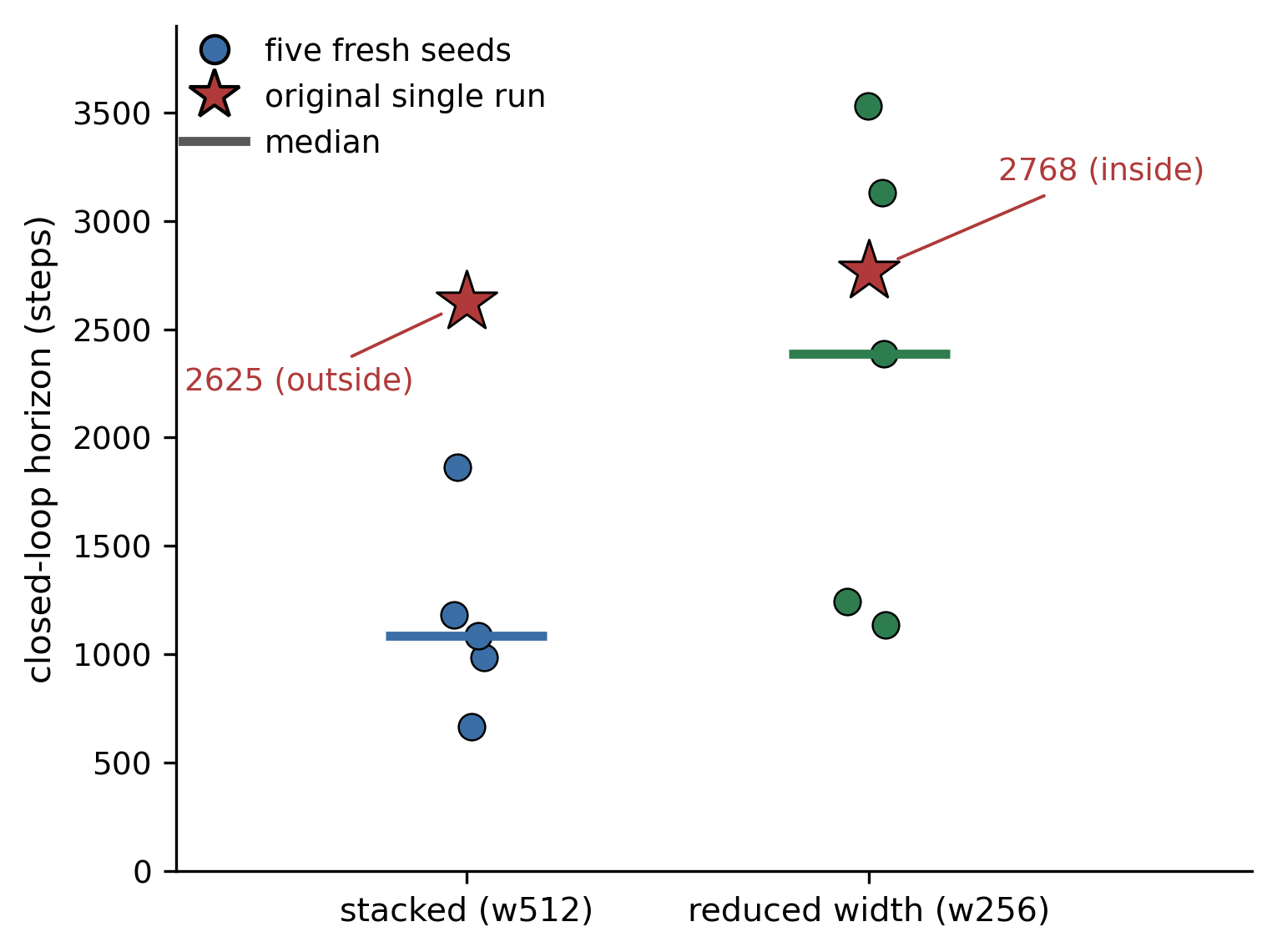}
\caption{Closed-loop horizon for five independently seeded trainings of the
stacked configuration and of its reduced-width counterpart, against the
single-run figures quoted elsewhere in this paper (stars). The original
$2625$-step figure falls outside its own five-seed distribution; the
reduced-width single run ($2768$) falls inside its own.}
\label{fig:n5repro}
\end{figure}

Single-run horizons of this kind are not reliable evidence about stability,
and a later $n=5$ check of this exact configuration shows why. Five freshly
trained seeds survive $666$, $985$, $1086$, $1183$ and $1862$ steps (median
$1086$), so the $2625$-step figure lies outside the entire distribution,
above every one of the five fresh runs (Fig.~\ref{fig:n5repro}). Data,
recipe and export pipeline were verified identical between the original run
and the check, so the most likely explanation is an unusually favourable
initialisation that was never independently reproduced rather than any
identified difference in method. We report $2625$ as the original verified
measurement but read it as a likely favourable outcome, with the five-seed
distribution as the more representative estimate of what this architecture
and recipe deliver. The same check covers the reduced-width network of
Sec.~\ref{sec:distill}, whose five seeds survive $1134$, $1243$, $2388$,
$3128$ and $3532$ steps (median $2388$); a Mann-Whitney test between the two
arms gives $U=3$, $p=0.056$.

\subsection{A regime-transfer warning}\label{sec:regime}

Aggregation is the most effective of the four levers pre-bounce. It does not
transfer to the post-bounce regime, and the direction of the failure is
worth reporting precisely because it is counterintuitive.

Applying the identical recipe post-bounce, five independently initialised
models were each trained on a baseline dataset and again after aggregation
augmentation (each augmented set combining the full prior archive with fresh
self-visited states, $3.5$ to $3.6$M records against the baseline's $2.3$M),
then evaluated by deterministic closed-loop rollout. All five got
\emph{worse}: mean survival fell from $64$ steps to $6$, a reduction of
$3.9$ to $19\times$ in every one of the five pairs, with genuine physical
divergence (the electron fraction oscillating between its guard floor and
ceiling within two to three steps) rather than a training artefact.

Aggregating states from short, already-unstable trajectories and retraining
pulls the network toward correcting those specific off-manifold points at
the expense of general behaviour near the training manifold, sharpening the
on-trajectory limit rather than loosening it. The same lever helping in one
regime and actively hurting in another is itself the lesson: no
stabilisation in this class transfers across regimes by default, and each
must be validated in the regime it is meant to fix.

\subsection{Stability is attainable; it is not the same as speedup}

Two results turn this into a constructive statement. A per-cell gate,
deferring only individual cells whose Mahalanobis distance exceeds a
threshold, stabilises the loop over the full test window (more than $1000$ steps with
no crash), so closed-loop stability \emph{is} attainable. And tracing the loop
against the true trajectory rather than counting steps to a crash yields a
physically motivated remedy: in the early-infall phase the true change in
$Y_e$ is essentially zero, so the surrogate's tiny but systematic error
accumulates ballistically, driving $Y_e$ from its correct, nearly frozen
$0.443$ down to $0.384$, whereupon the softened equation of state contracts
the core too fast. Freezing every fed-back channel where the coupling is
physically negligible, so that the surrogate is the identity exactly where
the true change is, reproduces the true central-density trajectory to six
figures through twelve thousand steps.

Both results are genuine and neither delivers acceleration. The freeze
enforces the identity precisely where the surrogate would otherwise have
been called, so it saves nothing; the substantive test is deferred to the
high-density phase where the freeze lifts. The gate is the subject of the
next section.

Table~\ref{tab:scoreboard} collects the horizons.

\begin{table}[!htbp]
\centering
\caption{Closed-loop horizon by method: steps the live simulation survives
before a genuine crash, with the resulting speedup or solver-deferral
fraction. Horizons are not directly comparable across rows, as they probe
different phases and stabilisations; all are in the early-infall regime. The
un-aided baseline for the ``$\times$'' factors is the thermodynamic-energy
model without drift coverage ($102$ steps).}
\label{tab:scoreboard}
\footnotesize
\setlength{\tabcolsep}{4pt}
\begin{tabular}{p{3.15cm} l l >{\raggedright\arraybackslash}p{3.75cm}}
\toprule
Configuration & Horizon (steps) & Speedup / defer & Note \\
\midrule
Naive surrogate (predicts $\tau,m_1$ too) & $\sim$2 & n/a & acute sub-resolution energy collapse ($100\times$ density) \\
\quad $+$ thermodynamic energy & 102 & pure replacement & acute obstacle removed; chronic $Y_e$ drift remains \\
Aggregation, naive iteration & $102$--$286$ (oscillates) & pure replacement & does not converge; last model $\neq$ best \\
Aggregation, keep-best $+$ recency & 557 ($5.5\times$) & pure replacement & $Y_e$ physical; delays but does not remove the crash \\
\quad $+$ Jacobian contractivity & $2625^{\dagger}$ ($26\times$) & pure replacement & levers stack; contractivity alone gives $\sim$36\% \\
Per-cell Mahalanobis gate & $>1000$ & $95$--$99\%$ solver & stable, but $\approx$0.95$\times$ once gate cost is included \\
Ensemble-disagreement gate & 1365 & $94\%$ to solver & better detector, still no speedup \\
Near-identity freeze ($Y_e$ only) & $\sim$5000 & pure replacement & central density tracks truth within $\sim$15\% \\
Near-identity freeze (all channels) & $\ge$12000 & pure replacement & reproduces central density to six figures \\
\midrule
Target: full collapse to bounce & $53{,}203$ & n/a & high-density phase is the substantive test \\
\bottomrule
\end{tabular}\\[2pt]
{\raggedright\footnotesize $^{\dagger}$A subsequent $n=5$ check of this
configuration (Sec.~\ref{sec:stabilise}) found this single-run figure lies
outside the resulting five-seed distribution (median $1086$, range
$666$--$1862$); read it as a verified but likely favourable outlier. \par}
\end{table}

\section{The economics of uncertainty gating}\label{sec:gate}

The uncertainty gate deserves separate treatment because it is the standard
remedy for exactly the failure diagnosed above, because its economics can be
written down in closed form, and because the result is the most restrictive
of the paper.

\subsection{Break-even in closed form}

Let $f$ be the target block's share of critical-rank wall clock, $d$ the
fraction of cells the gate defers to the exact solver, $r=t_{\rm
nn}/t_{\rm solver}$ the surrogate's per-call cost ratio, and
$g=t_{\rm gate}/t_{\rm solver}$ the gate's own per-cell cost ratio. Every
cell pays the gate; deferred cells then pay the solver, the remainder pay
the network. The end-to-end speedup is
\begin{equation}
S(d) \;=\; \frac{1}{1 - f\,\bigl[\,1 - \bigl(g + d + (1-d)\,r\bigr)\bigr]}\,,
\label{eq:gate}
\end{equation}
and the gate breaks even ($S=1$) at
\begin{equation}
d_{\rm break} \;=\; \frac{1-g-r}{1-r}\,.
\label{eq:breakeven}
\end{equation}
Two properties are worth noting. The break-even deferral is independent of
$f$: the runtime share sets how much a gate can win or lose, not whether it
wins. And the gate can never amortise once $g>1-r$, which for our measured
$r=0.17$ means $g>0.83$.

\subsection{Measured inputs}

We measured $g$ in situ at three implementation qualities. At the production
compiler settings actually deployed (plain \texttt{-O2}, no fast-math) the
full-covariance Mahalanobis test costs $g=0.391$ of a solver call, giving
$d_{\rm break}=0.53$. Optimised compiler flags roughly halve this to
$g=0.153$ ($d_{\rm break}=0.82$), and a diagonal covariance approximation
lowers it to $g=0.016$ ($d_{\rm break}=0.98$).

We flag one methodological trap encountered here, because it is easy to
repeat. Our first estimate of $g$ came from a standalone microbenchmark
compiled with aggressive flags and gave $g=0.031$, a factor $12.7$ below the
in-situ value. The gap decomposes into roughly $4.5\times$ from compiler
flags (the production translation unit is built at plain \texttt{-O2}),
$1.08\times$ from a standardisation loop the microbenchmark omitted, and the
remainder from in-situ cache effects. A component benchmarked outside the
build and call context of its deployment can misstate its cost by an order
of magnitude.

\subsection{Correct detection forces near-total deferral}

\begin{figure}[t]\centering
\includegraphics[width=\textwidth]{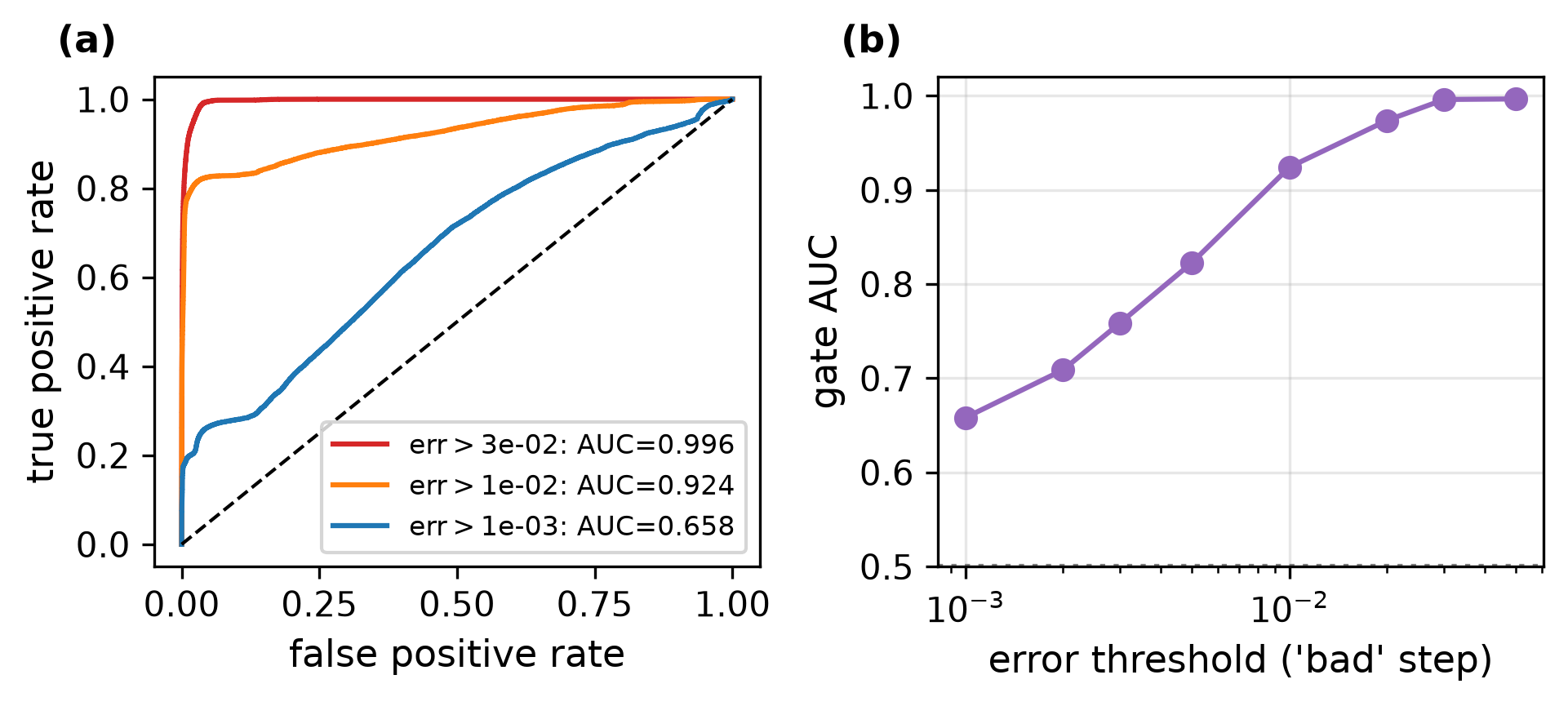}
\caption{Detection quality of the Mahalanobis gate. \emph{(a)} ROC curves
for three definitions of a ``bad'' step: the gate separates large blow-up
errors cleanly (err $>3\times10^{-2}$: AUC $0.996$) but the small
accumulating errors that constitute the drift only weakly (err
$>10^{-3}$: AUC $0.658$). \emph{(b)} Gate AUC as a function of the error
threshold, rising from $0.66$ to $0.996$: the detector is sharp exactly
for the failures that are not the binding ones.}
\label{fig:gate}
\end{figure}

The gate is a \emph{correct} out-of-distribution detector
(Fig.~\ref{fig:gate}): it separates
large blow-up errors cleanly (area under the curve $0.996$) while flagging
the small accumulating drift errors only weakly ($0.66$). In the loop it
prevents the crash. Precisely because it detects correctly, it routes almost
every cell to the solver: the measured deferral fraction has median $96.8\%$
(interquartile spread within a few tenths of a per cent; the $77.3\%$
minimum occurs only in the startup transient), so the network evaluates
$3.2\%$ of cells.

Substituting the deployed $g=0.391$, $d=0.968$, $r=0.17$ and the pre-bounce
critical-rank share $f=0.119$ (Table~\ref{tab:profile}) into
Eq.~\eqref{eq:gate}, the coupling block costs $136\%$ of its original
unguarded time and the end-to-end change is $0.96\times$: a measured
slowdown, not merely an absence of gain. The post-bounce share, $f=0.169$,
gives a slightly worse $S=0.94\times$ at the same operating point.

\subsection{Deferral is nearly invariant to surrogate quality}

The natural response is that a better surrogate would defer less. We
measured this directly, and it is the decisive result.

Two outer-zone models differing fivefold in training depth and fourfold in
closed-loop survival (median $326$ against $82$ steps) defer, respectively,
$98.1\%$ and $99.7\%$ of cells: a difference of $1.6$ percentage points.
Broken down by rank, the model sensitivity is concentrated in the four
high-volume outer-zone ranks, which carry eight times the gate evaluations,
and even there the deeper model never defers below $91\%$. Of the four
low-volume dense-core ranks, three sit at exactly $1.000$ for both models and
the fourth differs by one percentage point.

The deferral fraction is set by how pervasively the state has left the
training manifold, not by the surrogate's accuracy. A direct test with a
better detector confirms the same asymmetry from the other side: an
ensemble-disagreement detector, which separates drifted from valid states
markedly better than the Mahalanobis distance (area under the curve $0.96$
against $0.91$), extends the stable horizon ($1365$ against $1038$ steps)
but still defers $94\%$ of cells.

\begin{figure}[t]\centering
\includegraphics[width=0.72\textwidth]{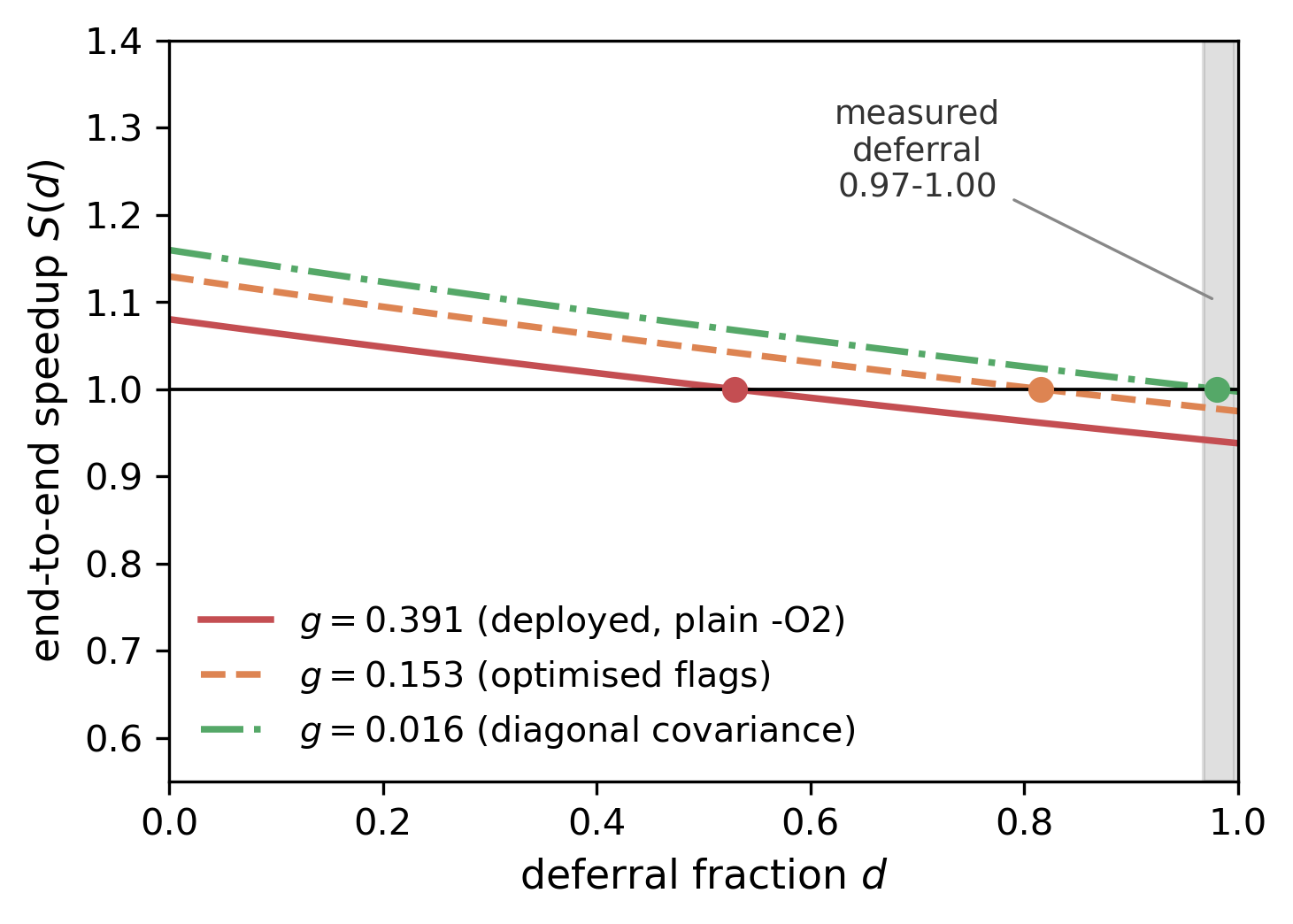}
\caption{Break-even deferral fraction as a function of gate cost, from
Eq.~\eqref{eq:breakeven}. Cutting the gate's cost by $24\times$ (deployed
$g=0.391$ to idealised diagonal $g=0.016$) only moves the break-even from
$0.53$ to $0.98$: the formula's dependence on $g$ is far shallower than its
dependence on $d$ near $d=1$. The measured deferral, $0.97$--$1.00$, sits
at or beyond all three.}
\label{fig:breakeven}
\end{figure}

The measured operating point, $d=0.98$ to $0.99$, sits at or beyond all
three break-evens including the idealised diagonal one (Fig.~\ref{fig:breakeven}). No gate
variant we tested, at any surrogate accuracy this project reached, yields a
speedup. The gate is not miscalibrated and it is not too expensive; a
correct out-of-distribution detector applied to a loop that leaves
distribution cannot accelerate it. To reach $d_{\rm break}=0.53$ from $0.98$
would require a surrogate orders of magnitude more accurate than anything
achieved here, and Sec.~\ref{sec:offline} shows we could not even reliably
\emph{select} such a model offline if we had it.

We verified both limits of the gate bitwise, since a mechanism this
consequential should not rest on an unverified switch: routing every cell to
the network reproduces the ungated network trajectory exactly, and routing
every cell to the solver reproduces a solver-only run exactly.

\subsection{A hybrid that localises the failure}

Because whole-network comparisons cannot say \emph{where} the loop fails, we
built a radial hybrid in which the network drives all cells denser than a
cutoff and the exact solver drives the rest, verifying both limits bitwise
as above.

The result separates two failure modes that every previous experiment had
conflated. Replacing the \emph{core} network by the exact solver, leaving
the network in the outer zone, reproduces the pure-network survival
distribution \emph{value for value across all five seeds}
($70,73,75,89,102$), with the same outer crash location: the specialist's
ceiling is set entirely by the outer zone, and its core network is never
driven to failure because the outer zone always fails first. Conversely,
replacing the \emph{outer} network by the exact solver raises survival to a
median of $108$ steps (against $75$; $p=0.024$) until an intrinsic
instability of the core network appears at $92$ to $131$ steps, now with no
seed reaching a healthy terminal state.

This also disposes of an intuition that is wrong in an instructive way. The
dense core is the stiffest region and the one where the surrogate
destabilises, so it is natural to assume it is where the solver spends its
time. It is not: under this cutoff the network drives only $12\%$ of cells,
and the outer ranks issue roughly $7.4\times$ more solver calls each. The
dense core is the \emph{unstable} region, not the \emph{expensive} one. A
hybrid offloading only the core therefore cannot save wall clock even if
perfectly stable, and on a call-count basis would cost more than deploying
the network everywhere.

Within this hybrid, matching the surrogate's training domain to the zone it
drives does help. Replacing the core specialist extrapolated onto the outer
zone with a network trained on the outer cells raises median survival from
$87$ to $224$ steps, with individual seeds reaching $868$. Tested on twelve
paired training seeds under a strictly reproducible protocol (verified by
retraining seeds to bitwise-identical weights), the magnitude-aware Wilcoxon
signed-rank test is significant ($W=6$, $p=0.007$), but the effect is not
uniform: three of twelve seeds favour the untargeted baseline, so the
magnitude-blind sign test is not significant ($9/12$, $p=0.15$), and an
eight-seed pilot showing seven of eight positive was optimistic in its
consistency. Training-domain match is a real and sizeable lever on
outer-zone survival, but a variable one across seeds. It changes nothing
about wall clock: network cost is architecture-bound and identical, the
mirror configuration is a measured $1.11\times$ slowdown, and a
longer-surviving surrogate keeps the expensive zone alive longer without
making the step faster.

\subsection{Pure replacement reaches parity and still fails}\label{sec:purerep}

The gate's alternative is pure replacement with no fallback. Here the cost
question has a subtlety that took us two rounds of measurement to see, and
we report both rounds because the difference between them is instructive.

The configuration stable over the early infall is a width-$512$ network of
$1.83\times10^{6}$ parameters. Evaluated through the per-cell scalar loops
in which it was first deployed, its forward pass costs
$2.3\times10^{-3}\,$s per cell against the solver's $3.4\times10^{-4}\,$s,
seven times more expensive than the routine it replaces, and the end-to-end
run is correspondingly $1.62\times$ slower than the pure solver ($232$
against $143\,$ms per step over a matched $2000$-step window). Evaluated
through a BLAS-backed path, the identical network with the identical weights
costs $2.7\times10^{-4}\,$s per cell, a factor of $8.3$ less, and $9.4
\times10^{-5}\,$s when the call is additionally batched across the cells of
one sweep, a further factor of $2.9$. The two implementations agree to
$6.7\times10^{-16}$ per call, at the floating-point reassociation floor.

Re-timed end-to-end on the BLAS path, with five runs per arm in alternating
order over a common window, the surrogate takes $213\pm13\,$ms per step
against the pure solver's $204\pm15$, a ratio of $0.96\times$ that a
permutation test cannot separate from parity ($p=0.41$). Absolute per-step
times are not comparable between the two rounds, which used different
windows under different node load, but the two are consistent where it
matters. The scalar arm's $89\,$ms per step of excess over its own baseline
is the difference between network and solver cost, so with the measured
per-call ratio of $6.6$ it implies a gross network cost of $105\,$ms against
a $16\,$ms solver block. Dividing the network term by $8.3$ predicts
$201\,$ms for the BLAS arm, against the $213\pm13$ observed: consistent
within the scatter, and predicting parity from the scalar-path numbers
alone. The
reduced-width
network of Sec.~\ref{sec:distill} reaches $192\,$ms per step on the same
path, nominally $1.06\times$, likewise inside the scatter.

The correction matters more than its size suggests. The cost penalty we
first measured, and the asymmetry we first drew from it, were properties of
an unoptimised inference path rather than of the surrogate. Once that is
fixed, pure replacement is cost-competitive: it neither pays a penalty nor
delivers a saving, which is exactly what the budget of
Sec.~\ref{sec:budget} predicts for a block at $11.9\%$ to $17.7\%$ of
critical-rank wall clock. The runs timed above use the unbatched BLAS path,
where the surrogate is already $1.3\times$ cheaper per call than the routine
it replaces; batching the call across a sweep would make it $3.6\times$
cheaper and still leave the end-to-end figure below the
$\approx\!1.2\times$ ceiling of Sec.~\ref{sec:budget}. That is the Amdahl argument in its strongest form, and it
does not depend on our network being slow.

What pure replacement fails on is therefore not cost. It is that the
configuration is demonstrated stable only over a few thousand of the
$53{,}203$ steps the infall phase takes, and that it is unfaithful within
that window (Sec.~\ref{sec:distill}). A cost-competitive surrogate that
cannot complete the phase, and drifts while it runs, is not a usable
replacement.

Running the same network through two implementations also produced a result
we did not anticipate, and it bears on how closed-loop horizons should be
reported. The two paths agree to $6.7\times10^{-16}$ per call, yet the
scalar-path run passes $2000$ steps while the BLAS-path run terminates at
step $293$ through the same causality guard that ends every other run in
this study. Neither is stochastic: four repetitions of the BLAS build give
$293$ steps every time, and its trajectory is reproducible digit for digit.
The horizon is thus a reproducible property of a network \emph{together
with} the arithmetic that evaluates it, not of the trained network alone.
Rounding differences at the reassociation floor are amplified by the
feedback loop into a factor of seven in survival. This sharpens the
five-seed result of Sec.~\ref{sec:distill}, where the spread could still be
attributed to different initialisations: here the weights are bit-identical.
A closed-loop horizon quoted without its inference path is therefore not a
reproducible number, and comparisons of horizons across papers, or across
builds within one project, should be read with that in mind.

\subsection{Can the stable network be compressed?}\label{sec:distill}

The asymmetry just described invites an obvious response: obtain the large
network's stability in a small, cheap one. We tested this, because it is the
natural next question. A width-$256$ network ($5.2\times10^{5}$ parameters,
a $3.5\times$ reduction and the same size as the cheap network that was not
stable on its own) was retrained from scratch on identical data with the
identical recipe that made the large one stable, aggregation data plus the
Jacobian-contractivity penalty, on $8.7\times10^{6}$ records. The large
network's outputs never entered its loss, so this is retraining at reduced
width rather than knowledge distillation. Against a pure-solver baseline of
$289$ and $288\,$s over the matched window, the large network takes
$467\,$s ($0.618\times$) and the reduced-width one $330$ and $327\,$s
($0.878\times$). Both figures are from the scalar inference path; on the
BLAS path of Sec.~\ref{sec:purerep} the two arms are $0.96\times$ and
$1.06\times$, so the cost gap between them is itself largely an artefact of
that path and the reduction buys little once inference is optimised. The $n=5$ check reported in Sec.~\ref{sec:stabilise} finds
no statistically demonstrable stability cost to the reduction, and if
anything the trend runs opposite to the naive expectation that the smaller
network would be less robust. Compressing further does not continue to help:
a width-$128$ network ($1.6\times10^{5}$ parameters) trained identically
crashed at step $1772$ and did not complete the matched timing window, so no
cost comparison could be obtained at that size. That last figure rests on a
single run and falls inside the five-seed range of the larger arm, so it
bounds the useful compression range only weakly.

Two conclusions follow. First, compression moves the result toward the
Amdahl ceiling; it does not move the ceiling. With the block between
$11.9\%$ and $17.7\%$ of critical-rank wall clock, even a \emph{free}
surrogate would cap between $1.14\times$ and $1.21\times$, and both arms
measured $0.96\times$ and $1.06\times$ on the BLAS path, inside run-to-run
scatter of parity and far below that cap. Second, compression does not
improve fidelity, and fidelity was poor in both arms from the outset.
Deleptonisation sets in as soon as the collapse starts and lowers $Y_e$
throughout, accelerating above
$\approx\!10^{11}\,\mathrm{g\,cm^{-3}}$; over the matched window analysed
here, which stays near $6\times10^{9}\,\mathrm{g\,cm^{-3}}$, that decline is
still slow enough that the reference $Y_e$ holds at $0.442630$ to the six
digits logged. Both networks depart from it within the first
hundred steps, both run low in entropy throughout, and the cheaper network
ends $24\%$ low in central density against the larger one's $3.3\%$. This
is the same dissociation between survival and fidelity documented below for
the gate in Sec.~\ref{sec:fidelity}, reached by an entirely different
route. A compression study evaluated on step count alone would have reported
a clear success.

The three stable configurations therefore span, in measured wall clock:
$0.94$ to $0.96\times$ (gated, including its detection cost), $0.96\times$ to
$1.06\times$ (pure replacement at full and reduced width on the BLAS
inference path, both inside run-to-run scatter), and parity within
contention noise (batched post-bounce surrogate, $0.296$ against
$0.318\,$s per step). None accelerates the code measurably, and the budget
of Sec.~\ref{sec:budget} says none could have exceeded
$\approx\!1.2\times$.

\section{Stability is not fidelity}\label{sec:fidelity}

A gated loop that never crashes is often reported as a success. Crash-freedom
and trajectory fidelity are different claims, and the speedup analysis above
does not settle the second. We tested it directly.

From an identical verified starting state we ran a matched pair: a gated run
deferring $99.88\%$ of cells to the solver and evaluating the network on
$0.12\%$, against a solver-only reference, both to $6000$ steps in the
pre-bounce regime.

\begin{figure}[t]\centering
\includegraphics[width=0.72\textwidth]{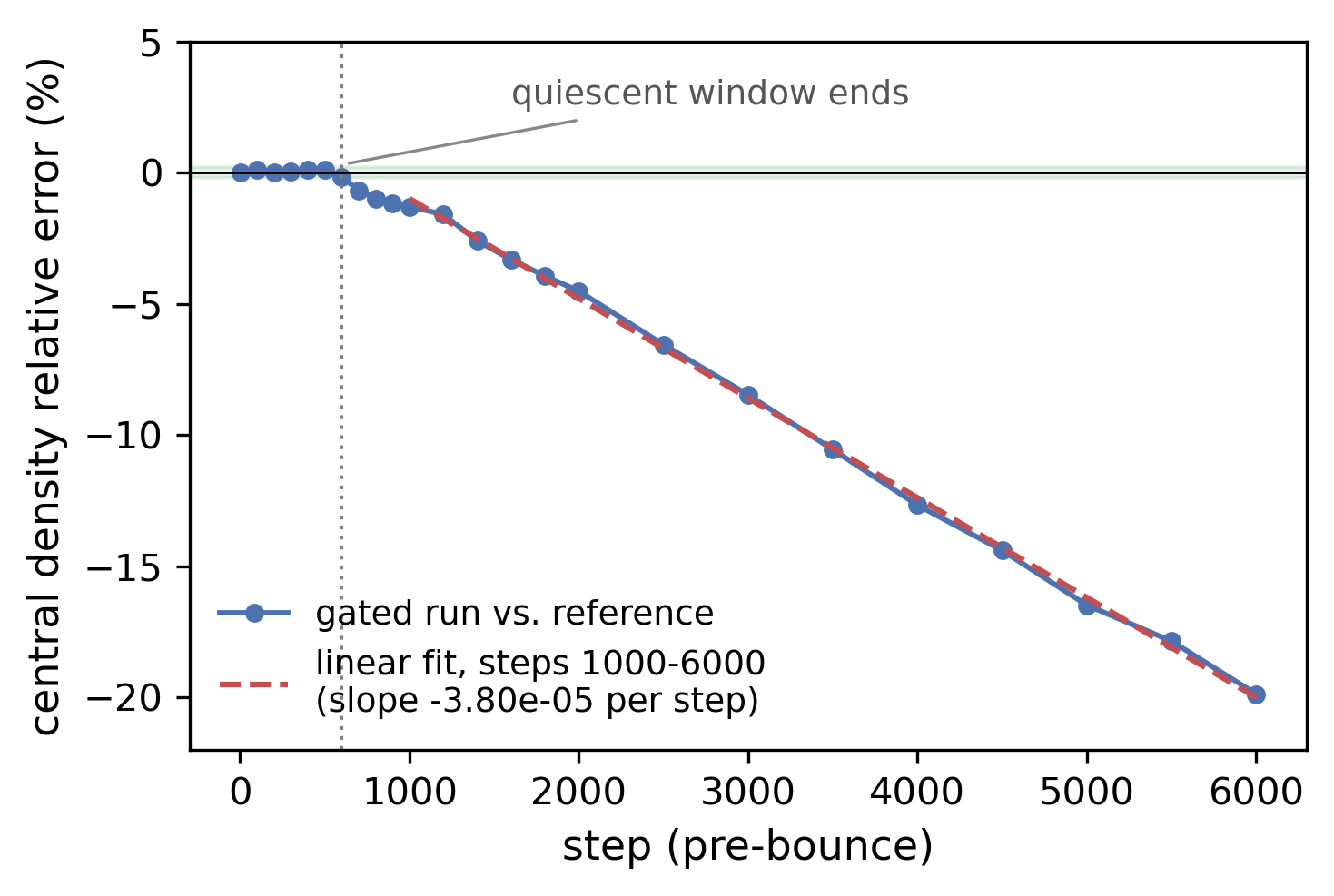}
\caption{Central density of a gated run relative to a solver-only reference
from the same starting state, pre-bounce. After a quiescent window of
$\sim\!600$ steps the error grows linearly (fitted slopes of $-3.6$ to
$-3.7\times10^{-5}$ per step over steps $1000$--$3000$, $3000$--$6000$, and
the full window, $R^2>0.993$ in each case), reaching $-19.9\%$ by step $6000$ despite
$99.88\%$ of cells being handled by the exact solver throughout.}
\label{fig:gatedrift}
\end{figure}

The electron fraction and entropy at a sampled cell jump to a small fixed
offset within the first $10$ to $25$ steps and then stay flat for the
remaining $\sim\!5975$ steps: no continued drift there. Central density
behaves differently. Its relative error stays under $0.2\%$ for roughly the
first $609$ steps, then grows \emph{linearly}, with a fitted slope between
$-3.6\times10^{-5}$ and $-3.7\times10^{-5}$ per step whether measured over
steps $1000$ to $3000$, $3000$ to $6000$, or the full window ($R^2>0.993$ in
each case), reaching $-19.9\%$ by step $6000$ (Fig.~\ref{fig:gatedrift}).

Three points follow. The growth is linear rather than accelerating, which is
consistent with the directed-bias mechanism of Sec.~\ref{sec:mechanism}
rather than with a variance-driven instability. It is
\emph{non-self-correcting}: deferring all but a fraction of a per cent of
cells to the exact solver does not pull the trajectory back. And it is
invisible to the crash criterion, so a stability-only evaluation would have
reported this run as a success.

We are explicit about scope. This test covers the pre-bounce regime, where
the gate's deferral fraction is at its most conservative, and the code's
fixed benchmark length was reached well before bounce. Whether the drift
stays linear, accelerates, or saturates in the post-bounce regime is not
established by this measurement and remains open.

\section{Mechanism: a directed bias, not a variance blow-up}\label{sec:mechanism}

A single causal chain organises the study, and identifying it correctly
determines which remedies can work.

The per-step error of the surrogate is small but \emph{systematic}: a
directed bias rather than zero-mean noise. A directed error accumulates
\emph{ballistically}, linearly in the number of steps, carrying the state
progressively off the training manifold; the network, accurate only on that
manifold, grows less accurate there, which accelerates the departure. Once
off-manifold the visited states are pervasively out of distribution, so any
correct gate must defer almost every cell and cannot accelerate. The
directed-bias measurement and the covariate-shift measurement are therefore
not competing diagnoses but the \emph{cause} and \emph{effect} of one
mechanism.

This distinguishes our setting from the standard framing. The
autoregressive-surrogate literature typically treats closed-loop instability
as \emph{variance-driven}, to be damped by input noise
\citep{sanchez2020gns} or a pushforward stability loss
\citep{brandstetter2022mppde}. Those remedies attack a fluctuation that is
here sub-dominant to a systematic drift.

\subsection{The bias is robust}

\begin{figure}[t]\centering
\includegraphics[width=0.72\textwidth]{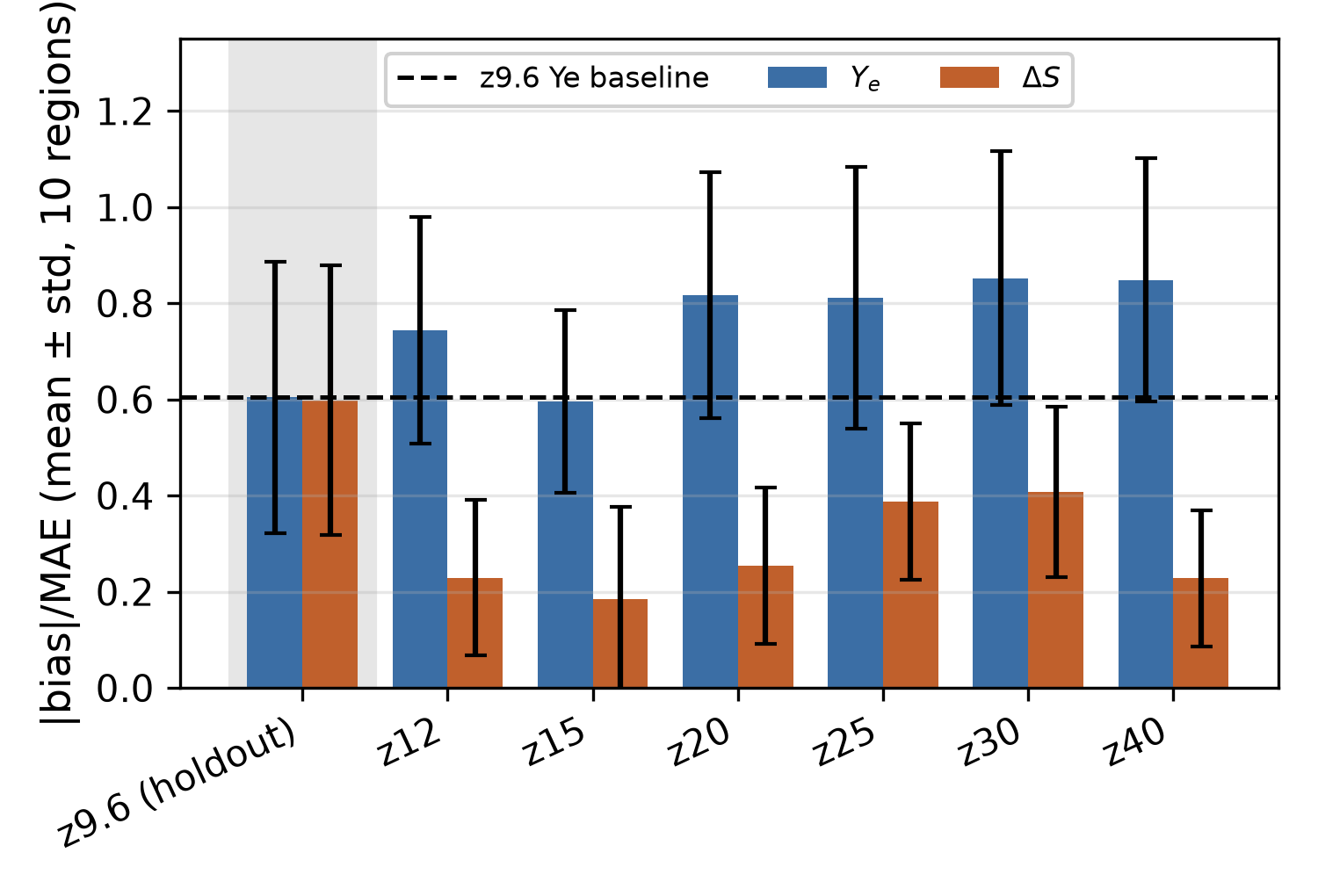}
\caption{The directed bias generalises across progenitors. The frozen model
is evaluated on the $20\,$ms post-bounce data of six independent progenitor
collapses ($12$ to $40\,M_\odot$); the directed electron-fraction bias
$|\mathrm{bias}|/\mathrm{MAE}$ stays at $0.60$ to $0.85$, comparable to the
$0.60$ baseline (dashed), while the entropy-increment bias is smaller on the
independent masses. Bootstrap over ten temporal regions.}
\label{fig:crossprog}
\end{figure}

We measured $|\mathrm{bias}|/\mathrm{MAE}=0.249\pm0.186$ post-bounce, robust
under bootstrap over three seeds and ten temporal regions. It is not an
artefact of the single progenitor: evaluating the frozen model on the full
$20\,$ms post-bounce data of six independent progenitors ($12$ to
$40\,M_\odot$), each an entirely separate collapse, the directed
electron-fraction bias persists at comparable magnitude
($|\mathrm{bias}|/\mathrm{MAE}$ of $0.60$ to $0.85$ against $0.60$ on the
trained holdout; Fig.~\ref{fig:crossprog}). Nor is it specific to one
regime: networks trained and evaluated on pre-bounce records show a directed
bias of the same size ($0.33\pm0.20$ for $Y_e$, $0.40\pm0.28$ for the
entropy increment), and a larger residual network offers no improvement
($0.40\pm0.22$).

\subsection{A systematic exclusion of remedies}

Against a directed, ballistically accumulating bias, none of the cheap
levers works. Not network capacity or architecture. Not training-set volume.
Not training-data diversity: in a controlled comparison at matched sample
size, a model trained on a mixture of five progenitor masses shows the same
directed bias as the single-progenitor model on every evaluation set,
including on masses contained in its own training mixture ($0.44$ to $0.53$
against $0.39$ to $0.54$, all within bootstrap error bars). And not offset
or conservation debiasing: decomposing the correction into a channel-wise
constant offset and a state-dependent part is structurally informative (the
entropy bias is almost entirely a constant offset, the electron-fraction
bias almost entirely state-dependent), but quantitatively negative. On the
$Y_e$ channel $|\mathrm{bias}|/\mathrm{MAE}$ falls only from $0.601\pm0.282$
to $0.497\pm0.346$ under a global scalar offset and $0.456\pm0.250$ under
the state-dependent correction, reductions lying entirely within the
bootstrap error bars. About three-quarters of the directed bias is
orthogonal to any offset- or conservation-based correction, so a
conserved-quantity correction of the kind that stabilises autoregressive
neural operators elsewhere \citep{current2026cqc} does not transfer to this
regime.

Two levers of a different kind remain, and both follow from the mechanism
rather than from generic regularisation. The first exploits a structural
property of a directed bias: it has a roughly fixed absolute floor while the
true per-step physical change shrinks with the timestep. Predicting over the
physical timescale $\tau_\varepsilon=|X|/|\mathrm{d}X/\mathrm{d}t|$ and
rescaling the increment by $\mathrm{d}t/\tau_\varepsilon$ shrinks the
delivered per-step bias by $\langle\mathrm{d}t/\tau_\varepsilon\rangle$
without touching the physical part. The ratio of bias to physical change
drops by one to two orders of magnitude in every physically active regime
(for $Y_e$ from $0.092$ to $0.0046$, a factor of $20$; for the entropy
increment from $0.080$ to $0.0016$, a factor of $49$), and it transfers to all six independent progenitors ($Y_e$:
$0.016$ to $0.032$; entropy: $0.0006$ to $0.0015$). The second is
off-manifold coverage, which the covariate-shift analysis independently
identifies, and which has also been reported as decisive for a closely
analogous surrogate of stiff nuclear burning
\citep{grichener2025nnn}.

The per-step figures above are the direct measurement, and for a rescaling
lever they predict the reduced ballistic accumulation, because ballistic
growth is by construction linear in the per-step bias. In the loop the
picture is more restrictive, and we report it rather than resting on the
per-step proxy. Applied on its own post-bounce, the rescaling extends
survival from $50$ to $59$ steps, an $18\%$ gain that is of no practical
use. The reason is that it addresses the wrong failure mode: a channel
ablation localises the acute post-bounce death to the radiation-moment
prediction, not to $Y_e$ and entropy. Freezing the radiation moments to
their incoming values alone carries the loop to $2582$ steps, and adding
the timescale rescaling on top of that carries it to at least $9998$ steps,
the point at which the run terminated cleanly at its $2\,$ms cutoff rather
than crashing. The lever therefore does contribute in the loop, but only
once the dominant failure mode has been removed by other means, and the
combined configuration is a diagnostic rather than a usable scheme: it
survives, but its central density settles at
$4.6\times10^{11}\,\mathrm{g\,cm^{-3}}$ instead of returning to
proto-neutron-star values, so it is another instance of survival without
fidelity rather than a remedy.

\subsection{Where pushforward training does and does not help}

The regime-dependence of Sec.~\ref{sec:regime} recurs here, in the opposite
direction, which strengthens rather than weakens the point.

Pushforward training \citep{brandstetter2022mppde} made no measurable
difference pre-bounce, where the drift is a covariate shift into states no
single-trajectory curriculum covers. Post-bounce the failure has a different
character, and there it works offline: over an eight-step rollout the plain
model's radiation-channel error grows roughly fortyfold, the ballistic
accumulation expected of the feedback loop, whereas the pushforward-trained
model's error stays flat, at the cost of a few-fold worse one-step accuracy.
This is the first lever in this study to measurably suppress rollout
accumulation offline.

The closed loop bears out both the promise and the caveat. Substituting the
pushforward-trained model into the same restart test, changing only the
weights, the run survives $65$ steps rather than $31$, and the specific
dense-core instability diagnosed there does not recur at the tracked cell.
But the same guard eventually fires, this time at a low-density outer cell
far from the two dense-core cells the pushforward training saw. The
regularisation transfers partially and genuinely, doubling survival and
removing the failure mode it was trained against, but it relocates rather
than eliminates the instability, moving it to a region outside its training
diversity, exactly as the two-cell caveat predicts.

\subsection{Attribution controls: solver-only and subcycled runs}

One control deserves emphasis because it is the kind that is easy to omit.
To establish that the instabilities above are caused by the substitution
rather than by the host code in this regime, we ran the unmodified solver
alone through the same window, bypassing the network entirely at every step.
It completes the full $20\,$ms window without incident, with central density
flat within $0.1\%$ of nuclear saturation over its final $3000$ steps, in
clear contrast to every network-driven run. The instability is attributable
to the substitution, not to the grid, domain size, or a latent numerical
issue.

A second control cuts the other way and is equally necessary. A subcycling
scheme calling the solver only every $N$-th step, swept over
$N\in\{1,5,10,20\}$, produced an unexpected result. In all four arms the
network drives the electron fraction and entropy at every step, so none is a
solver-only reference; the sweep varies only how often the radiation is
solved exactly. $N=1$, the arm that solves it every step, is the
\emph{only} configuration that aborts, while
$N{=}5,10,20$ all reach the cutoff. They do so, however, at central
densities of order $10^{5}$ to $10^{6}\,\mathrm{g\,cm^{-3}}$, eight to nine
orders of magnitude below nuclear saturation and five to six below the
$8.6\times10^{11}\,\mathrm{g\,cm^{-3}}$ at which the reference failed. Every subcycled configuration reaches an unphysical
end state; the only distinction is whether the numerical degeneracy crosses
a floating-point exception threshold before the cutoff. Subcycling converts
a hard crash into a silently unphysical but numerically stable trajectory,
which is a worse failure mode for any application that trusts a
non-aborting run, and we do not report it as a usable operating point.
Measured wall clock across the four runs ($0.270$, $0.279$, $0.276$,
$0.256\,$s per step) shows no discernible speedup at any $N$ in any case.

\section{Discussion and transferable guidance}\label{sec:guidance}

\subsection{What generalises}

Our result is not that neural surrogates fail, nor that this particular
network was inadequate. It is that three specific quantities determine
whether a surrogate for an inner solver block can accelerate a simulation,
and that all three can be measured before substantial modelling effort is
spent.

\paragraph{The runtime share bounds everything} Per-call cost is the
quantity that is easy to measure and the one most often reported; the
validated share of critical-path wall clock is the quantity that bounds the
achievable gain. Our target block was the most expensive routine per call in
its code and still only $16.9\%$ of critical-rank wall clock. A surrogate
$5.8\times$ cheaper per call tied.

\paragraph{Offline metrics do not rank deployments} Across fourteen models
the only significant pooled correlation was a between-family confound, and
within families the metric had no resolution at all. If model selection
matters, it must be done in the loop, on distributions.

\paragraph{The economics of gating are unfavourable}
Equation~\eqref{eq:breakeven} can be evaluated before a gate is built, from
a per-call cost ratio and an estimate of the deferral fraction. The
counterintuitive part is that deferral is set by the physics of the drift
rather than by the detector or the surrogate: ours moved $1.6$ percentage
points across a fourfold change in survival.

\paragraph{Stability and fidelity are different acceptance criteria} A
gated run deferring $99.88\%$ of cells never crashed, yet drifted $-19.9\%$
in central density over $6000$ steps.

\paragraph{Identify the error's character before choosing a remedy} A
directed bias accumulates linearly and is attacked by contractivity,
timescale rescaling, or off-manifold coverage; a variance-driven blow-up is
attacked by noise injection or pushforward losses. Applying the second class
to the first damps a sub-dominant term, which is consistent with what we
measured.

\paragraph{Validate stabilisations in the regime they are meant to fix}
Dataset aggregation gave a $5.5\times$ horizon gain pre-bounce and a $3.9$
to $19\times$ degradation post-bounce, across five paired seeds.

\subsection{A pre-deployment checklist}

For practitioners considering a surrogate for an inner solver block, we
suggest the following order of operations, which is roughly the reverse of
the order we followed.

\begin{enumerate}
\item Profile the full time step with exclusive timers and validate the
decomposition by summation. Compute $1/(1-f)$ on the critical path. If that
ceiling does not justify the effort, stop here.
\item Measure the per-call cost of a surrogate of the size the problem
plausibly needs, in the build and call context of deployment rather than a
standalone benchmark.
\item Check whether the target increment is representable at all. If its
true magnitude is at or below the achievable regression error, do not
regress it; derive it from learnable quantities through the exact relations
already implemented.
\item Establish seed variance before comparing configurations, and compare
distributions.
\item If gating is contemplated, estimate $d$ from a short instrumented run
and evaluate Eq.~\eqref{eq:breakeven} before building the gate.
\item Evaluate fidelity against a reference trajectory, not only survival.
\item Re-validate every stabilisation separately in each regime of the
simulation.
\end{enumerate}

\subsection{Limitations}

We state the boundaries of these claims plainly. All closed-loop and timing
measurements are from a single code, a single progenitor, and a single node
configuration; only the bias-generalisation tests of
Sec.~\ref{sec:mechanism} use additional progenitors. The
runtime share is a property of this code and configuration, and while the
volume-scaling test found no path to a solver-dominated regime, we do not
extrapolate it to production scale. The load-imbalance estimate is a model
with stated bracketing assumptions, not a measurement. The fidelity test of
Sec.~\ref{sec:fidelity} covers only the pre-bounce regime. The compression
study of Sec.~\ref{sec:distill} rests on two timing runs per arm, so its
cost figures are solid, but on a single stability run per arm; the
five-seed follow-up reported in the same section has since replaced that
single-run horizon comparison with a distributional one; the width-$128$ result rests
on a single crashed run. The GPU analysis of Sec.~\ref{sec:gpu} is an analytical estimate,
not a measurement. And the
chronic drift is diagnosed precisely but not solved: every stable
configuration we tested is stable by deferring to the solver, by enforcing
the identity where the coupling is negligible, or over a horizon short of
the full collapse.

We also note what a favourable case this was. The target block has a
per-cell, state-in/state-out interface with no spatial stencil and no
history, the training data are abundant and exactly labelled by the solver
itself, and the surrogate reaches $R^2=0.98$ on held-out data. A block with
a less convenient interface would not have done better on any of the three
barriers.

\section{Conclusions}\label{sec:conclusion}

We set out to accelerate a stiff implicit coupling solve with a neural
surrogate, in a setting deliberately chosen to be favourable: the target is
the most expensive routine per call in its code, its interface is per-cell
and history-free, and exact training labels are abundant. No configuration
we tested accelerates the simulation end-to-end, and the reasons are
structural.

The block is $16.9\%$ of critical-rank wall clock, capping any surrogate at
$\approx\!1.2\times$; a surrogate $5.8\times$ cheaper per call ties the
solver, and the surrogate stable enough to run unaided is $1.3\times$
cheaper per call on the inference path we timed, $3.6\times$ cheaper when
the call is batched, and at parity end-to-end ($0.96\times$, $p=0.41$). Offline accuracy
cannot select among surrogates: the pooled correlation across fourteen
models is a between-family confound and vanishes ($\rho=-0.04$) under
control. A correct out-of-distribution gate must defer $96.8\%$ to $99.7\%$
of cells because the loop leaves its training distribution within a few
steps, a fraction nearly invariant to surrogate quality, placing the
operating point beyond every break-even we measured; including its own cost
the gated loop is a $0.94$ to $0.96\times$ slowdown. Crash-freedom does not imply
fidelity: a gated run deferring $99.88\%$ of cells accumulates a linear
$-19.9\%$ central-density bias over $6000$ steps.

Two results are constructive. When a target increment lies below the
achievable regression error, predicting the learnable thermodynamic state
and deriving the rest through the exact algebraic relations removes an acute
instability outright ($2\to102$ steps). And closed-loop stability is
attainable: dataset aggregation with best-model retention and a recency
window, stacked with a Jacobian-contractivity penalty, reaches a median of
$1086$ closed-loop steps across five independently seeded trainings (one
verified single run reached $2625$, $26\times$ the un-aided baseline, though
that figure lies outside the five-seed distribution and should be read as a
favourable outlier rather than typical) as a pure replacement. Neither
delivers acceleration, and the second holds only in the regime it was
validated in: the same aggregation degrades post-bounce survival by $3.9$
to $19\times$.

The credible route to a real speedup in this block is not a learned
surrogate but a restructured exact solver, for which
\citet{laiu2020thornado} report up to $100\times$ on GPU for this same
physics, and, in our code, the static-decomposition load imbalance that
dominates the profile and is not a machine-learning problem at all.

We regard the methodology as the transferable contribution: measure the
runtime share before the per-call cost, validate in the loop on
distributions rather than single runs, evaluate gate economics against the
break-even fraction before building the gate, and treat stability and
fidelity as separate acceptance criteria. Above all, a surrogate for a
stiff, coupled simulation is trustworthy only on-trajectory, and per-step
accuracy alone can be confidently, and dangerously, misleading.

\section*{CRediT authorship contribution statement}
\textbf{L.~Th\"ummler:} Conceptualization, Methodology, Software,
Validation, Formal analysis, Investigation, Data curation, Writing --
original draft, Writing -- review \& editing, Visualization.
\textbf{T.~Kuroda:} Software (original radiation-hydrodynamics code),
Resources, Writing -- review \& editing.

\section*{Declaration of competing interest}
The authors declare that they have no known competing financial interests
or personal relationships that could have appeared to influence the work
reported in this paper.

\section*{Funding}
This research did not receive any specific grant from funding agencies in
the public, commercial, or not-for-profit sectors.

\section*{Data and code availability}
The instrumented solver, the surrogate training and inference code, and the
logged coupling datasets are available from the authors on reasonable
request.

\section*{Acknowledgements}
Computations were performed on the yamazaki cluster at the Max Planck
Institute for Gravitational Physics (Albert Einstein Institute).

\section*{Declaration of generative AI and AI-assisted technologies in the
manuscript preparation process}
During the preparation of this work the authors used Anthropic Claude in
order to translate author-written text into English. After using this tool,
the authors reviewed and edited the content as needed and take full
responsibility for the content of the published article.

\end{document}